\documentclass[a4paper,11pt]{article}
\pdfoutput=1 

\usepackage{jheppub} 

\newcommand{\cR}{\mathcal{R}}

\def\be{\begin{equation}}
\def\ee{\end{equation}}
\def\beqn{\begin{eqnarray}}
\def\eeqn{\end{eqnarray}}

\def\ba{\begin{array}{c}}
\def\bat{\begin{array}{cc}}
\def\ea{\end{array}}
\def\bi{\begin{itemize}}
\def\ei{\end{itemize}}

\def\be{\begin{equation}}
\def\ee{\end{equation}}

\title{\boldmath Fingerprinting the presence of extra scalars in the $h \to Vff'$ forward-backward asymmetry.}

\author[a]{Victor Ilisie}
\emailAdd{victor.ilisie@i3m.upv.es}

\affiliation[a]{I3M, Universitat de Polit\`ecnica Val\`encia - CSIC, Camino de Vera s/n, 46022, Valencia, Spain}

\abstract{In this work we analyse the forward-backward asymmetry of the $h\to Vff'$ decay in the Aligned two-Higgs Doublet Model. The Standard Model prediction for this asymmetry for $V=W$ is small, as it suffers from Yukawa suppression and is absent for $V=Z$. This does not necessarily have to hold true in the Aligned model where these contributions can in principle be re-enhanced through the independent alignment factors $\varsigma_f$. In this analysis we conclude that, due to the additional contributions corresponding to the Aligned two-Higgs Doublet Model together with extra sources of CP-violation for the $V=Z$ channel, the Standard Model predictions can be significantly modified in a great region of the parameter space. These deviations, that could be potentially measured at the High Luminosity LHC or future Higgs factories, would be a clear signal of new physics, and would shed new light on the possible extensions of the Standard Model and new sources of CP-violation.}

\begin{document}

\maketitle
\flushbottom

\section{Introduction}
\label{sec:introd}

The discovery of a Higgs-like scalar boson by the ATLAS \cite{Aad:2012tfa} and the CMS \cite{Chatrchyan:2012xdj} collaborations has opened the possibility to look for new physics effects, specially by analysing its different decay channels. Even though, the latest provided data seem to be amazingly consistent with the Standard Model (SM) \cite{Aad:2019mbh,Cadamuro:2019tcf,ATLAS-CONF-2019-004, Aaboud:2018urx,CMS-PAS-HIG-19-005,Sirunyan:2018koj,Sirunyan:2018kst}, as some of the experimental uncertainties are still rather large, we can use it as a great source to further constrain many SM extensions. In this work we will analyse the Aligned Two-Higgs Doublet Model (ATHDM) extension of the SM, with exclusive interest in the $\varphi_i^0\to Vff'$ channels, where $V$ is a gauge boson and $\varphi_i^0$ is one of the neutral scalars of the model which we identify with the discovered SM-like boson with $M_{\varphi_i^0} = 125$ GeV.

Currently the only experimentally available LHC data for these channels is provided for the $V=W$ with $(f,f')=(l,\nu_l)$, where the on-shell $W$ boson subsequently decays into a second pair of lepton-neutrino \cite{Aaboud:2018jqu,Aad:2019lpq,Sirunyan:2020tzo} and, $V=Z$ with $(f,f')=(l^+, l^-)$, where similarly, the on-shell $Z$ boson decays into a second pair of leptons \cite{ATLAS:2020wny,Aaboud:2017oem,Sirunyan:2017tqd,Sirunyan:2017exp,Sirunyan:2018sgc}. Instead of the leptonic channel, one could also focus on light quark decays i.e., $(f,f') = (q,q')$ (that hadronize to light mesons). However, the corresponding experimental signals of such decays are much more challenging to single out, as the background is considerably larger, dominated by QCD processes. Also, the only potential significant contribution (due to the mass suppression) would correspond to a final state bottom quark. However, any kinematically allowed decay to a bottom quark and an up-type quark for the $V=W$ channel (which excludes the top quark) will be suppressed by the non-diagonal CKM matrix elements. Therefore, in this study we shall exclusively focus on final state leptons for $V=W$. For $V=Z$ we will study both final state leptons and quarks and the reinterpretation of the corresponding equations (when switching from leptons to quarks) is trivial.

\begin{figure}[!t]
\centering
\includegraphics[scale=0.45]{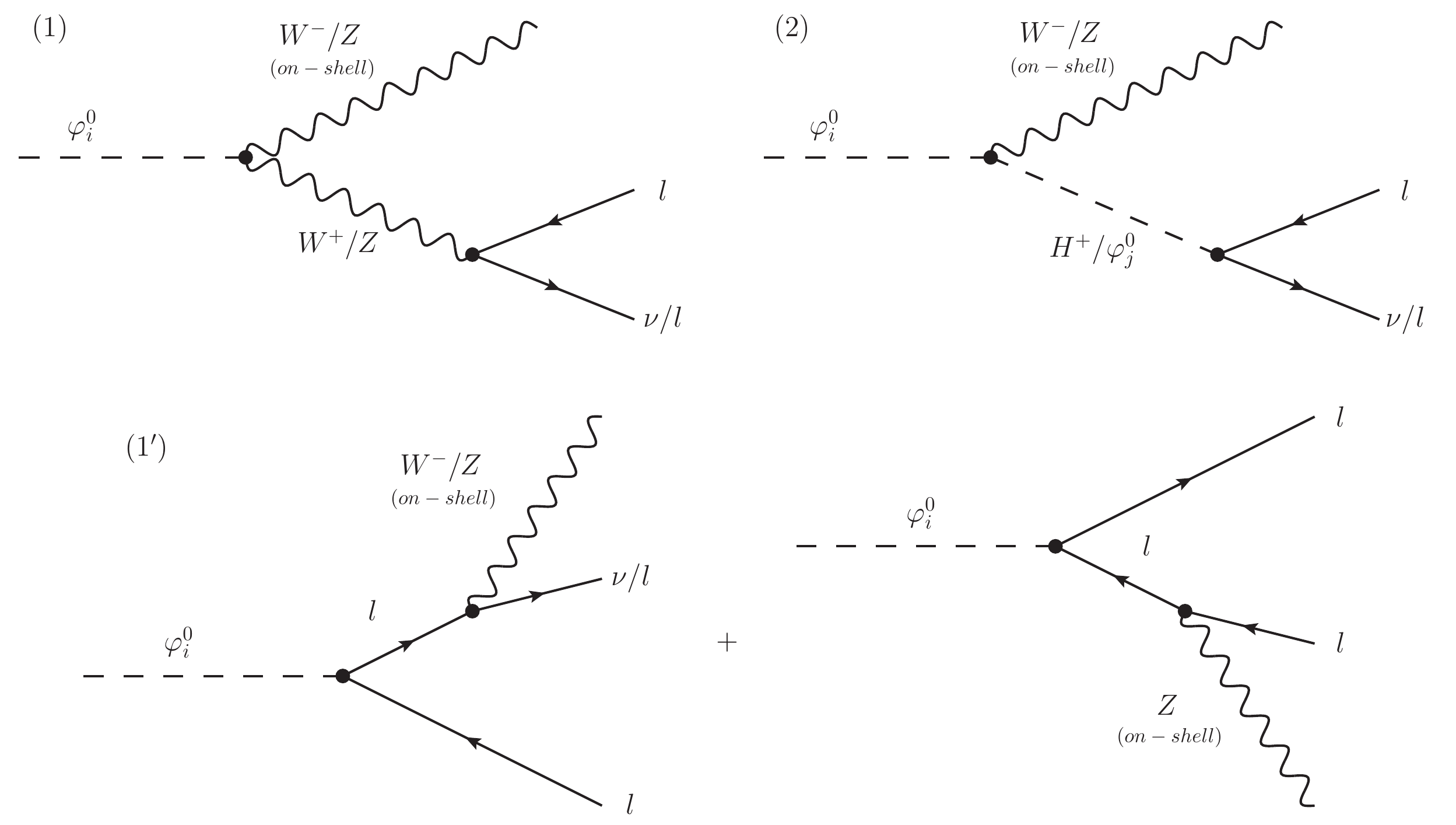}
\caption{Feynman diagrams for the $\varphi_i^0\to W^-l^+\nu$ and $\varphi_i^0\to Z^-l^+l^-$ decay.}
\label{diagramsh4f}
\end{figure}

The SM contributions that account for the three-body decay are shown in Fig.~\ref{diagramsh4f}, diagrams $(1)$ and $(1')$, where $\varphi_i^0$ stands for the SM-like Higgs boson. The extra charged/neutral Higgs contributions from the ATHDM are given by diagram $(2)$. For the SM calculations, for the total decay rates, one normally ignores diagram(s) $(1')$. However, as the FB asymmetry is proportional to the Yukawa couplings (terms which can be ignored for the total widths up to a good approximation), one has to necessarily include these contributions in order to obtain the correct complete expression for the FB asymmetry. In fact we shall see that, roughly speaking, the contributions to the FB asymmetry of $(1')$ and $(2)$ have opposite signs and $(1')$ dominates over $(2)$ for the $V=W$ case. 

However, the most outstanding prediction corresponds to the $V=Z$ case for which the SM prediction for the FB asymmetry is zero. We shall see that, by allowing extra sources of CP-violation, we can obtain important deviations from this prediction in the ATHDM. Thus measuring an angular asymmetry in this channel would be a unmistakable signal of new physics.

If we express the differential decay rate for the previously mentioned processes in terms of the $ff'$ invariant mass $q^2$, and the polar angle $\theta$ defined defined in the $ff'$ center of mass (CM) frame (see Appendix~\ref{AppA} for the conventions used in this analysis) we can define the forward-backward (FB) asymmetry  
\begin{align}
\mathcal{A}_{\text{FB}}^V = \left[ \left( \int_{-1}^0  - \int_0^1 \right) d\cos\theta \; \frac{d^2\Gamma(\varphi_i^0\to Vff')}{dq^2 \, d\cos\theta} \right] \bigg{/} \frac{d^2\Gamma(\varphi_i^0\to Vff')}{dq^2}  \, .
\label{AFBV}
\end{align}
Measuring the FB asymmetry might result rather challenging when final state $\tau$-leptons are involved. Similar analyses have been performed in a rather different context such as $B\to D^{(*)}\tau\nu$ decays \cite{Bhattacharya:2020lfm, Asadi:2020fdo, Hill:2019zja, Tanaka:1994ay, Sakaki:2013bfa, KEUNE,Datta:2012qk, Celis:2012dk, Fajfer:2012vx, Hu:2018veh, Murgui:2019czp, Alonso:2017ktd, Cheung:2020sbq}. Several methods have been proposed since, in order to reconstruct the $\tau$-lepton momentum (which is needed for extracting the polar angle) \cite{Bhattacharya:2020lfm, Asadi:2020fdo, Hill:2019zja, Tanaka:1994ay, Sakaki:2013bfa, KEUNE} however such a measurement has not yet been performed.

The SM predicts a small FB asymmetry for this decay as it is proportional to the Yukawa couplings ($m_f/v$). In the ATHDM one can re-enhance these contributions through the aligned factors $\varsigma_f$ and therefore, obtain predictions of this observable, that can  substantially deviate from the SM prediction in many regions of the parameter space. Particularly, for the $V=Z$ channel, in order to obtain a non-zero asymmetry one has to consider extra CP-violating effects in either the potential, the Yukawa sector or both. In this analysis we shall focus on CP-violating effects in the Yukawa sector, i.e., complex alignment coefficient $\varsigma_l$, and a CP-conserving potential for both channels ($V=W, \,Z$). We shall also briefly comment on the case corresponding to real Yukawa parameters and small CP-violating effects in the scalar potential in Appendix~\ref{CP-pot}.

Any future experimental data on this observable would shed new light on possible SM extensions and possible new sources of CP-violation. With all this being said, LHC data are already available for the differential cross section corresponding to the $4l$ final state \cite{ATLAS:2020wny,Sirunyan:2017tqd} in effective field theory (EFT) frameworks \cite{Gonzalez-Alonso:2014eva, Greljo:2015sla}. Even if we cannot reinterpret the EFT Higgs pseudo-observables in terms of the ATHDM parameters in order to obtain applicable constraints to our model, as we cannot match our results onto the $hJ^\mu_f J^\nu_{f'}g_{\mu\nu}$ effective operator (where $J^\mu_f \sim \bar{f}\gamma^\mu f$), the results are extremely promising.\footnote{Similar theoretical studies in terms of dimension 6 operators of the $hVff'$ coupling have also been performed in \cite{Isidori:2013cla, Pomaroll, Banerjee:2019pks, Falkowski:2015jaa}.}

The impact of the total decay rate (including the contributions of the extra scalars) on the measured LHC signals strengths is given by the quantity
\begin{align}
{\mu}_V = \frac{\sigma(pp\to \varphi_i^0X) \times \text{BR}(\varphi_i^0 \to V ff')}{\sigma^{\text{SM}}(pp\to h X) \times \text{BR}^{\text{SM}}(h\to V ff')} \, ,
\label{mu_V}
\end{align}
where $\sigma(pp\to \varphi_i^0X)$ stands for any generic production channel of the $\varphi_i^0$ boson. In Appendix~\ref{sign}, we derive the generalized signal strengths, that will be used in this work in order to obtain the relevant experimental constraints applicable to our analysis. We shall observe that, including the extra contributions has little effect on the signal strengths, however this conclusion cannot be extrapolated to the FB asymmetry. This should result clear from the following. If one ignores diagram(s) $(1')$ from Fig.~\ref{diagramsh4f}, then the double differential decay rate can be expressed as
\begin{align}
\frac{d^2\Gamma(\varphi_i^0\to Vff')}{dq^2 \, d\cos\theta} = a^V(q^2) - b^V(q^2) \cos \theta + c^V(q^2) \cos^2 \theta \, ,
\label{diffDecay}
\end{align}
and the FB asymmetry takes the form
\begin{align}
\mathcal{A}_{\text{FB}}^V = \frac{b^V}{2 \, a^V + \dfrac{2}{3} \, c^V} \, .
\end{align}
Thus, one should note that after integrating (\ref{diffDecay}) on $\cos\theta$ in order to obtain the total decay rate, the contribution of $b^V$ vanishes. Consequently, the total width encodes no information on the $b^V$ coefficient thus, on the FB asymmetry.

In this work we first briefly remind the basic aspects of the ATHDM. Afterwards we present the formulae for angular coefficients. Using the generalized LHC signal strengths (Appendix~\ref{sign}) we perform a fit to the LHC experimental data for a scenario corresponding to a CP-conserving potential and CP-violating Yukawa couplings. We then proceed with the numerical analysis of the given observables and finally we present our conclusions. As we shall observe, the simple three-body kinematics corresponding to these two channels, can provide a clear signal of new physics, as the FB asymmetry can be substantially modified by the presence of the extra scalars and additional sources of CP-violation, when compared to the SM predictions.

Some additional comments are required. For this analysis we have a very limited parameter space namely the set  $(\varsigma_l, \, \tilde{\alpha}, \, M_{H^\pm})$ or the set $(\varsigma_d, \, \tilde{\alpha}, \, M_{H^\pm})$ (depending if we look at lepton or quark final states). These three parameters (corresponding to each independent case\footnote{We will not be concerned in this analysis about possible correlations between the FB asymmetry in the lepton and the quark sector, which would imply the study of correlations between $\varsigma_l$ and $\varsigma_d$.}) do not exhibit any relevant joint correlation in the flavour sector, muon anomalous magnetic moment or electric dipole moments (EDMs) \cite{
Eberhardt:2020dat, Abbas:2018pfp, Hu:2016gpe, Celis:2012dk, Abbas:2015cua, Ilisie:2015tra, Celis:2013ixa, Jung:2010ab, Jung:2010ik, Celis:2013jha, Jung:2013hka} (where most stringent experimental bounds come from). The significant bounds are always expressed in terms of products of two alignment factors $\varsigma_f^* \, \varsigma_{f'}$ (with $f\neq f'$), or in terms of products involving some additional triple scalar coupling i.e., $\varsigma_f \, \lambda_{\varphi_i^o H^+ H^-}$, in expressions that can additionally involve scalar masses $M_{\varphi_i^0}$ or $M_{H^\pm}$. Therefore, for the allowed regions of the ($\varsigma_l, \, \tilde{\alpha}, \, M_{H^\pm}$) or ($\varsigma_d, \, \tilde{\alpha}, \, M_{H^\pm}$) parameter space satisfying the LHC data only, we will always be able to find regions for the remaining parameters of the model, for which the additional constraints (from the flavour sector, EDMs, etc.) are satisfied. This is why, in our current study we will only use the LHC experimental data.


\section{The Aligned Two-Higgs-Doublet Model}
\label{sec:A2HDM}

The 2HDM extends the SM with a second scalar doublet of hypercharge $Y=\frac{1}{2}$. In the Higgs basis 
$(\Phi_1,\Phi_2)$, only one doublet acquires a vacuum expectation value:
\begin{equation}  \label{Higgsbasis}
\Phi_1=\left[ \begin{array}{c} G^+ \\ \frac{1}{\sqrt{2}}\, (v+S_1+iG^0) \end{array} \right] \; ,
\qquad\qquad\qquad
\Phi_2 = \left[ \begin{array}{c} H^+ \\ \frac{1}{\sqrt{2}}\, (S_2+iS_3)   \end{array}\right] \; ,
\end{equation}
with $G^\pm$ and $G^0$, the Goldstone fields.
In this basis $\Phi_1$ plays the role of the SM scalar doublet with
$v = (\sqrt{2}\, G_F)^{-1/2} = 246~\mathrm{GeV}$.
The physical scalar spectrum contains two charged fields $H^\pm(x)$
and three neutral scalars $\varphi_i^0(x)=\{h(x),H(x),A(x)\}$. They are related to the original $S_i$ fields
through an orthogonal rotation $\varphi^0_i(x)=\mathcal{R}_{ij} S_j(x)$.
The $\mathcal{R}$ matrix is fixed by the scalar potential. A detailed discussion can be found in \cite{Celis:2013rcs}. For the most generic (CP-violating) potential, the CP-odd component $S_3$ mixes with the CP-even fields $S_{1,2}$ and the resulting mass eigenstates do not have a definite CP quantum number.

The most generic Yukawa Lagrangian with the SM fermionic content gives rise to FCNCs because the fermionic couplings of the two scalar doublets cannot be simultaneously diagonalized in flavour space. These FCNCs can be either kept under control (BGL models \cite{Botella:2015hoa}), eliminated through an appropriate ${\cal Z}_2$ symmetry \cite{Glashow:1976nt} or by requiring the alignment in flavour space of the Yukawa matrices~\cite{Pich:2009sp}; {\it i.e.}, the two Yukawa matrices coupling to a given type of right-handed fermions are assumed to be proportional to each other and can, therefore, be diagonalized simultaneously. This will give rise to three proportionality parameters $\varsigma_f$~($f=u,d,l$) which are arbitrary complex numbers and introduce new sources of CP violation.

In terms of the fermion mass-eigenstate fields, the Yukawa interactions of the A2HDM read~\cite{Pich:2009sp}
\beqn\label{lagrangianYuk}
 \mathcal L_Y & = &  - \frac{\sqrt{2}}{v}\; H^+ \left\{ \bar{u} \left[ \varsigma_d\, V M_d \mathcal P_R - \varsigma_u\, M_u^\dagger V \mathcal P_L \right]  d\, + \, \varsigma_l\, \bar{\nu} M_l \mathcal P_R l \right\}
\nonumber \\
& & -\,\frac{1}{v}\; \sum_{\varphi^0_i, f}\, y^{\varphi^0_i}_f\, \varphi^0_i  \; \left[\bar{f}\,  M_f \mathcal P_R  f\right]
\;  + \;\mathrm{h.c.} \, ,
\eeqn
where $\mathcal P_{R,L}\equiv \frac{1\pm \gamma_5}{2}$ are the right-handed and left-handed chirality projectors,
$M_f$ the diagonal fermion mass matrices
and the  couplings of the neutral scalar fields are given by:
\begin{equation}    \label{yukascal}
y_{d,l}^{\varphi^0_i} = \cR_{i1} + (\cR_{i2} + i\,\cR_{i3})\,\varsigma_{d,l}  \, ,
\qquad\qquad
y_u^{\varphi^0_i} = \cR_{i1} + (\cR_{i2} -i\,\cR_{i3}) \,\varsigma_{u}^* \, .
\end{equation}
The usual models with natural flavour conservation, based on discrete ${\cal Z}_2$ symmetries, are recovered for particular (real) values of the couplings $\varsigma_f$ \cite{Pich:2009sp}. The coupling of a single neutral scalar with a pair of gauge bosons takes the form ($V=W,Z$)
\begin{align}
g_{\varphi_i^0 VV} = \mathcal{R}_{i1} \; g^{\text{SM}}_{hVV}\, ,
\label{sumrule}
\end{align}
which implies $g_{hVV}^2 + g_{HVV}^2 + g_{AVV}^2 = (g_{hVV}^\text{SM})^2$. Thus, the strength of the SM Higgs interaction is shared by the three 2HDM neutral bosons. In the CP-conserving limit, the CP-odd field decouples while the strength of the $h$ and $H$ interactions is governed by the corresponding $\cos\tilde\alpha$ and $\sin\tilde\alpha$ factors.

\section{Differential Decay Widths}

In the following we shall present the results corresponding to the differential decay rate and the FB asymmetry for the two cases, as a function of the leptonic invariant mass $q^2$ (with $\sqrt{q^2} \in [m_f+ m_f', \, M_{\varphi_i^0} - M_W]$) and the polar $\theta$ angle defined in the leptonic CM frame (see Appendix~\ref{AppA} for a brief overview). For our analysis, it will be useful to express the double differential decay rate as
\begin{align}
\frac{d^2\Gamma(\varphi_i^0\to Vff')}{dq^2 \, d\cos\theta} = a^V(q^2) - b^V(q^2) \cos \theta + c^V(q^2) \cos^2 \theta + \Lambda^V(q^2,\cos\theta) \, ,
\label{diffDecay_2}
\end{align}
where the contributions to the $\Lambda^V(q^2,\cos\theta)$ term are given by the squared matrix of diagram $(1')$ and the corresponding crossed terms.\footnote{Due to the fact that the fermion propagator contains terms proportional to $\cos\theta$ in the denominator, we cannot express this term as previously, as a linear combination of the type $a - b \cos\theta + c\cos\theta^2$ with $a,b$ and $c$ functions that do not depend on $\cos\theta$.} If we define
\begin{align}
\Lambda^V(q^2) \equiv \int_{-1}^1 d\cos\theta \,  \Lambda^V(q^2,\cos\theta) \, ,  \quad \Lambda_{\text{FB}}^V(q^2) \equiv  \left( \int_{-1}^0 - \int_{0}^1\right)  d\cos\theta \,  \Lambda^V(q^2,\cos\theta) \, ,
\label{lamb_fb}
\end{align}
then, the FB asymmetry takes the form
\begin{align}
\mathcal{A}_{\text{FB}}^V(q^2) = \frac{b^V(q^2) + \Lambda_{\text{FB}}^V(q^2) }{2 \, a^V(q^2) + \dfrac{2}{3} \, c^V(q^2) + \Lambda^V(q^2)} \, . 
\label{A_lambda}
\end{align}
Similar to the expression given in (\ref{diffDecay_2}), and because it will extremely useful to calculate the cancellations that occur for $V=Z$ for the FB asymmetry, we will also express $\Lambda^V(q^2,\cos\theta)$ as
\begin{align}
\Lambda^V(q^2,\cos\theta) = a^V_\Lambda(q^2,\cos\theta) - b^V_\Lambda(q^2,\cos\theta) \cos \theta + c^V_\Lambda(q^2,\cos\theta) \cos^2 \theta \, ,
\end{align}
where the coefficients $a^V_\Lambda, \, b^V_\Lambda$ and $c^V_\Lambda$, contain in their denominator terms that are proportional to $\cos\theta$.

We shall numerically analyse these results in Section~\ref{phen}, together with the relevant experimental constraints. In the following we are going to present the explicit formulae for the previously introduced terms. As we are interested to study the asymmetry for the whole range of $q^2$, we are going to keep all the terms proportional to $m_l$ that exhibit non-trivial 
behaviour for $q^2\simeq m_l^2$ i.e., of the type $m^2_l/q^2, \, m^4_l/q^4$, etc. Additionally, we have checked numerically that the terms that have been left out introduce no significant error. The complete formulae for the corresponding $\Gamma(\varphi_i^0 \to W^- l^+\nu)$ and $\Gamma(\varphi_i^0 \to Zl^+l^-)$ differential decay rates in terms of the previously introduced angular coefficients are given in Appendix~\ref{AppForm}.

\section{Phenomenology}
\label{phen}

A brief analysis of the previous results in terms of a CP-violating scalar potential and real Yukawa sector are given in Appendix~\ref{CP-pot}. In the following we shall focus our phenomenological analysis on the assumption that the scalar potential is CP-symmetric and that the Yukawa sector is CP-violating. Thus the scalar field mixing $\varphi_i^0 = \mathcal{R}_{ij} S_j$ is explicitly given by
\begin{align}
\begin{pmatrix}
h \\ H \\ A 
\end{pmatrix} 
=
\begin{pmatrix}
 \cos\tilde{\alpha} & \sin\tilde{\alpha} & 0 \\
-\sin\tilde{\alpha} & \cos\tilde{\alpha} & 0 \\
             0      & 0                  & 1
\end{pmatrix} 
\begin{pmatrix}
S_1 \\ S_2 \\ S_3 
\end{pmatrix} 
\, ,
\end{align}
where we identify $h$, $H$ as the CP-even scalars and $A$ with the CP-odd boson. We shall also assume that ${\varphi_{i=1}^0}=h$ is the lightest scalar of the model with $M_h = 125$ GeV. On the other hand we will consider a CP-violating Yukawa sector, which translates into complex values of $\varsigma_f$. Thus, the $y^h_f$ Yukawa couplings for $f=l,d$ (which are the only two Yukawa couplings we are interested in) will be given by 
\begin{align}
y^h_{l,d} &= \cos \tilde{\alpha} +  \sin \tilde{\alpha} \, \text{Re}\left(\varsigma_{l,d}\right) + i \, \sin \tilde{\alpha} \, \text{Im}\left(\varsigma_{l,d}\right) \, ,
\label{yuk_sig}
\end{align}
where we have separated $\varsigma_{l,d}$ as
\begin{align}
\varsigma_{l,d} = \text{Re}\left( \varsigma_{l,d} \right) + i \, \text{Im} \left(\varsigma_{l,d}\right) \, . 
\end{align}
The coupling of $h$ to a pair of gauge bosons, will be simply ($\kappa_V^{h}\equiv
g_{hVV}/g_{hVV}^{\mathrm{SM}}$, $V=W,Z$)
\be\label{equations2}
\kappa_V^{h}\;=\; \mathcal{R}_{11} \; = \; \cos{\tilde \alpha} \, . 
\ee

Note that for the $V=Z$ channel, we have only one extra contribution that corresponds to the CP-odd scalar boson $A$. Its leptonic Yukawa coupling takes the form
\begin{align}
y^A_l = -\text{Im} \left(\varsigma_l\right) + i \, \text{Re}\left( \varsigma_l \right).
\end{align}
A quick look at expression (\ref{B12}) reveals that only the real part of $y^A_l$ (thus the imaginary part of $\varsigma_l$) participates in this process. Also, when performing the integration on $\cos\theta$ in order to obtain $\Lambda^Z_{\text{FB}}(q^2)$ (see eq.~\ref{lamb_fb}), the only surviving terms (as we shall see in subsection~\ref{zff}) will be $\mathcal{C}^Z   \mathcal{A}_{1'2}^Z ( a_{1'2}^Z - b_{1'2}^Z \, \cos\theta)$. Again, by taking a quick look at the expressions of $a_{1'2}^Z$ and $b_{1'2}^Z$ in (\ref{a1p2}) we observe that they vanish in the CP-conserving limit, i.e., the combination of Yukawa couplings $\mathcal{Y}_{ij}^\text{R}$ and $\mathcal{Y}_{ij}^\text{I}$ that explicitly read (for $i=1$ and $j=3$, which is the only possibility in the CP-conserving limit of the potential)
\begin{align}
\mathcal{Y}_{ij}^\text{R} &= \text{Re}(y^h_{l,d}) \, \text{Re}(y^A_{l,d}) = - \big(\cos\tilde{\alpha} + \sin\tilde{\alpha} \; \text{Re} (\varsigma_{l,d}) \big) \, \text{Im}(\varsigma_{l,d}) \, , \notag \\[1.5ex]
\mathcal{Y}_{ij}^\text{I} &= \text{Im}(y^h_{l,d}) \, \text{Im}(y^A_{l,d}) = \sin\tilde{\alpha} \; \text{Im} (\varsigma_{l,d}) \, \text{Re}(\varsigma_{l,d}) \, ,
\label{Yuk_comb}
\end{align}
only survive for non-zero values of $\text{Im}(\varsigma_{l,d})$. Hence the need for CP-violating effects in the Yukawa sector in order to obtain non-zero contributions to the FB asymmetry in the ATHDM for the $V=Z$ channel.

As previously explained, due to the fact that we have a reduced parameter space $(\varsigma_l, \, \tilde{\alpha}, \, M_{H^\pm})$ or $(\varsigma_d, \, \tilde{\alpha}, \, M_{H^\pm})$, the allowed region for these parameters subject to the Higgs LHC experimental constraints will satisfy the extra bounds from the flavour sector, EDMs, etc., as the later ones depend on products of the type $\varsigma_f^* \, \varsigma_{f'}$ (with $f\neq f'$) or $\varsigma_f \, \lambda_{h H^+ H^-}$ for example (where additional couplings are involved). Therefore, in what follows, we shall perform a $\chi^2$ fit (with the expressions for the signal strengths presented in \cite{Celis:2013rcs} and the corresponding generalizations given in Appendix~\ref{sign})
\begin{align}
\chi^2 = \sum_j \left( \frac{\hat{\mu}_j - \mu_j}{\sigma_j} \right)^2 \, ,
\end{align} 
to the latest LHC Higgs data \cite{Aad:2019mbh, Cadamuro:2019tcf, ATLAS-CONF-2019-004, Aaboud:2018urx, CMS-PAS-HIG-19-005, Sirunyan:2018koj, Sirunyan:2018kst,Aaboud:2018jqu,Aad:2019lpq,Sirunyan:2020tzo,Sirunyan:2017exp,Sirunyan:2018sgc,ATLAS:2020wny,Aaboud:2017oem,Sirunyan:2017tqd} in terms of the real and imaginary parts of the Yukawa couplings i.e., in terms of $\text{Re}(y^h_f)$ and $\text{Im}(y^h_f)$ with $f=\{u,d,l\}$. 
Afterwards, from the leptonic Yukawa coupling in one case, and from the down-type Yukawa in the other case, we will extract the corresponding correlations between the real and imaginary parts of $\varsigma_l$ or $\varsigma_d$ and $\tilde{\alpha}$ by using (\ref{yuk_sig}).

Indirect bounds on $M_{H^\pm}$ or $M_A$ can be extracted from the extra contributions to the $h\to VV^*$ decay (Appendix ~\ref{sign}), however these constraints will turn out to be insignificant, as these extra contributions to the total decay widths are very suppressed. The $H^\pm$ also contributes to the $h\to \gamma\gamma$ decay, however as it also involves the coupling $\lambda_{hH^+H^-}$, we can always consider it small enough so that the experimental data are satisfied. 

\begin{figure}[!t]
\centering
\includegraphics[scale=0.38]{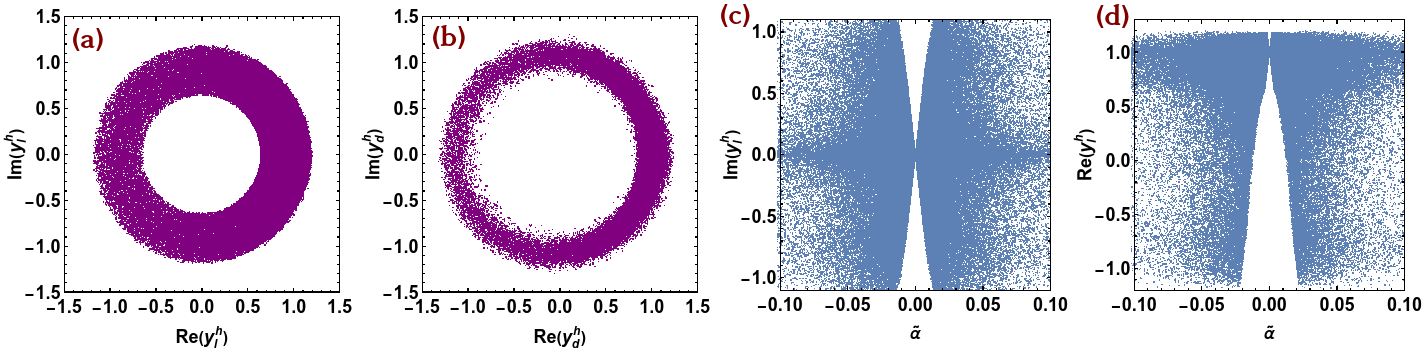} 
\caption{Allowed regions from global fits to the LHC data for Im$(y^h_l)$ $-$ Re$(y^h_l)$ (a), Im$(y^h_d)$ $-$ Re$(y^h_d)$ (b), Im$(y^h_l)$ $-$ $\tilde{\alpha}$ (c) and for Re$(y^h_l)$ $-$ 
$\tilde{\alpha}$ (d).}
\label{yhlconstr}
\end{figure}

In conclusion, we will allow the free parameters to vary in the following intervals (where the upper bounds of $\varsigma_{l,d}$ are given by perturbativity)
\begin{gather}
|\varsigma_l|  \in [0, \,100] \, ,  \qquad
|\varsigma_d|   \in [0,\, 50] \, , \qquad
\tilde{\alpha} \in  \left[-{\pi}/{2} , \, {\pi}/{2} \right]  \, , \notag \\[1.5ex]
 M_{H^\pm} \in [80, \,500] \, , \qquad \qquad M_A \in [50, \, 500] \, \, ,
\end{gather}
where $|\varsigma_{l,d}| = \left[ \text{Im}(\varsigma_{l,d})^2 + \text{Re}(\varsigma_{l,d})^2\right]^{1/2}$ and where the values of $M_{H^\pm}$ and $M_A$ are given in GeV. The obtained allowed regions for the (real and imaginary parts of) $y^h_l$ and $y^h_d$ are shown in Fig.~\ref{yhlconstr} (a) and (b). Correlations between Im$(y^h_l)$ and $\tilde{\alpha}$ and, Re$(y^h_l)$ and $\tilde{\alpha}$ are also shown i.e., Fig.~\ref{yhlconstr} (c) and (d). Similar plots can be obtained for Im$(y^h_d)$, Re$(y^h_d)$ and $\tilde{\alpha}$. 

\begin{figure}[!t]
\centering
\includegraphics[scale=0.38]{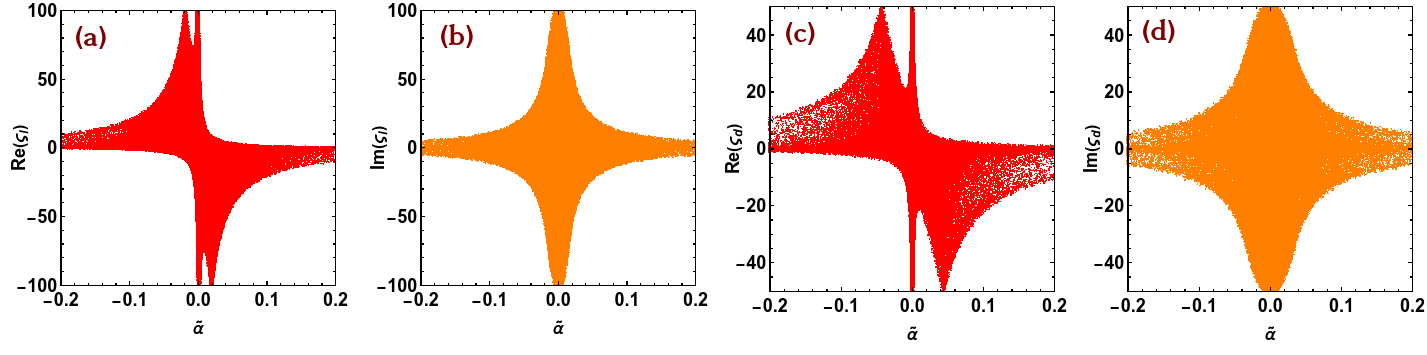} 
\caption{Allowed regions for Re$(\varsigma_l)$ and Im$(\varsigma_l)$ as functions of $\tilde{\alpha}$, shown in (a) and (b), and same for Re$(\varsigma_d)$ and Im$(\varsigma_d)$, shown in (c) and (d).}
\label{yhlconstr_2}
\end{figure}

Finally, if we reinterpret these previous constraints in terms of $\varsigma_l$ and $\varsigma_d$ we obtain the allowed regions shown in Fig.~\ref{yhlconstr_2}. No significant constraints are found for the masses $M_{H^\pm}$ and $M_A$ in terms of $\varsigma_{l}$, $\varsigma_{d}$ and $\tilde{\alpha}$. In the following we are going to use as an input the allowed regions for $(\varsigma_l,\, \tilde{\alpha}, \, M_{H^\pm})$ and $(\varsigma_d,\, \tilde{\alpha}, \, M_A)$ for the analysis of the total decay rates and the FB asymmetry.

\subsection{Total decay widths}

In this section we analyse the impact on the total decay width when including the extra contributions corresponding to diagrams $(1')$ and $(2)$. The total decay width, for a gauge boson $V$, is obtained simply by integrating on $\cos\theta$ and $q^2$ 
\begin{align}
\Gamma(h \to V ff') &= \int_{-1}^1 d\cos\theta \int_{q_0^2}^{q_1^2} dq^2 \Big( a^V(q^2) - b^V(q^2) \cos\theta +  c^V(q^2) \cos^2\theta   +  \Lambda^V(q^2,\cos\theta) \Big) \notag  \\
&= \int_{q_0^2}^{q_1^2} dq^2 \left( 2 \, a^V(q^2) + \frac{2}{3} \,  c^V(q^2) + \Lambda^V(q^2) \right) \, ,
\end{align}
where $q_0^2 = (m_f+m_f')^2 $ and $q_1^2 = (M_{h}-M_V)^2$.

In order to quantify the effect of the extra diagrams on the total decay rate we define the following quantity
\begin{align}
\omega^V_{\Gamma} = \frac{\Gamma(h\to Vff') - \Gamma^{(1)}(h\to Vff')}{\Gamma^{(1)}(h\to Vff')} \, ,
\end{align}
where $\Gamma(h\to Vff')$ includes the contributions from all diagrams and $\Gamma^{(1)}(h\to Vff')$ includes only the contributions from diagram $(1)$.

\begin{figure}[!t]
\centering
\includegraphics[scale=0.4]{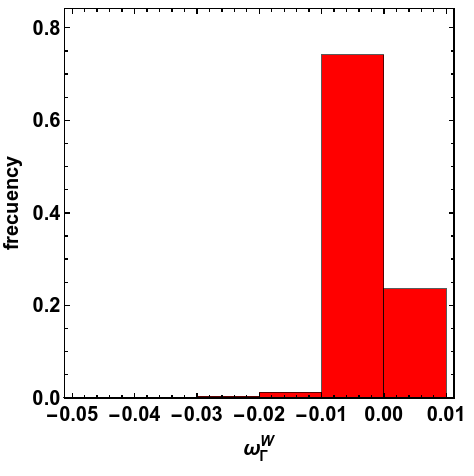} 
\caption{Values of $\omega_\Gamma^W$ and its frequency (normalized to 1) for a parameter scan in the allowed regions for $\tilde{\alpha}, \, \varsigma_l$ and $M_{H^\pm}$ for $(f,\,f') = (\tau, \, \nu_\tau)$ final state fermions.}
\label{W_Z_g}
\end{figure}

For $V=W$ with $(f,\,f') = (\tau, \, \nu_\tau)$ final state fermions we can observe in Fig.~\ref{W_Z_g} that the absolute values for $\omega_{\Gamma}^W$ are extremely small, of the order of mostly a few percent, which is far beyond the current experimental precision at the LHC. We can therefore safely conclude, that the total decay widths hardly notice the extra contributions, and the same conclusion can be extended for the signal strengths (\ref{mu_V}). For the remaining final state fermions and also for the $V=Z$ channel, the results are similar.

\subsection{FB asymmetry}

In order to quantify the impact on the FB asymmetry, that arises from the new degrees of freedom of the ATHDM when compared to the SM prediction, we define the following quantity for the $V=W^-$ channel
\begin{align}
\omega_{\rm{FB}}^W \equiv \frac{\mathcal{A}^W_{\rm{FB}}(q^2)- \mathcal{A}^W_{\rm{FB}}(q^2)_{\rm{SM}} }{\mathcal{A}^W_{\rm{FB}}(q^2)_{\rm{SM}}} \, ,
\end{align}
where the SM counterparts can be obtained from the corresponding expressions in the ATHDM taking the $\mathcal{R}_{i1}\to 1$ limit. 

On the other hand, for $V=Z$, things are a little different. As the SM prediction zero \cite{Isidori:2013cla}, we shall study the FB asymmetry without relating it to the SM. In the following we shall analyse $\omega_{\rm{FB}}^W$ for $(f,\,f') = (\tau^+, \, \nu_\tau)$ and, $\mathcal{A}_{\text{FB}}^Z$ for $(f,\,f') = \{(\tau^+, \, \tau^-), \, (b, \, \bar{b})\}$.

\subsubsection{$Wff'$ Channel}

\begin{figure}[!t]
\centering
\includegraphics[scale=0.44]{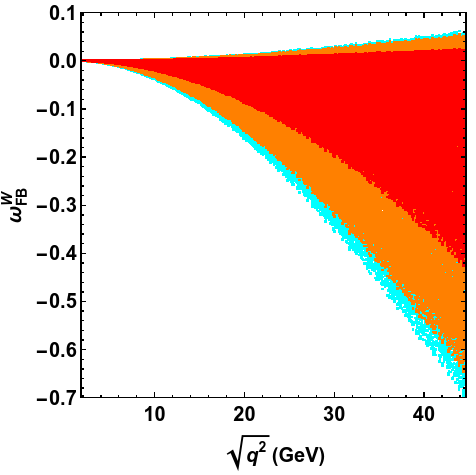} \quad  \includegraphics[scale=0.44]{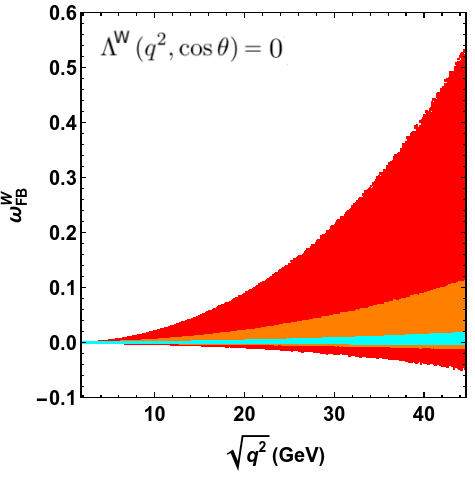}  
\caption{Left: Values of $\omega_{\rm{FB}}^W$ for $M_{H^\pm}$ 100 GeV (red narrow region),  200 GeV (orange wider region) and 500 GeV (cyan widest region). Right: values of 
$\omega_{\rm{FB}}^W$, setting $\Lambda^W(q^2,\cos\theta) = 0$ for the same values of $M_{H^\pm}$, 100 GeV (red wide region), 200 GeV (orange narrower region) and 500 GeV (cyan narrowest region).}
\label{yhlconstrW_a_mu}
\end{figure}

We will dedicate this section to the analysis of the $V=W^-$ channel with $(f,\,f') = (\tau^+, \, \nu_\tau)$ final state fermions. In Fig.~\ref{yhlconstrW_a_mu} (left) we show the results for $\omega_{\rm{FB}}^W$ for the whole range of $\sqrt{q^2}$ for different values of $M_{H^\pm}$ i.e., 100 GeV (red narrow region), $200$ GeV (orange intermediate region) and $500$ GeV (cyan widest region). We can observe that the ATHDM predictions can deviate from the SM up to almost 60-70\% for charged Higgs masses heavier than 200 GeV. At first sight one might find this result counter-intuitive, as one expects smaller contribution as $H^\pm$ becomes heavier. However, this only due to the fact that, as mentioned previously, the terms from $\Lambda^W_{\text{FB}}(q^2,\cos\theta)$ kinematically dominate over $b^W(q^2)$. In Fig.~\ref{yhlconstrW_a_mu} (right) we present the values of $\omega_{\rm{FB}}^W$ neglecting the contributions from diagram $(1')$ (by setting $\Lambda^W(q^2,\cos\theta) = 0$ in eq.~\ref{diffDecay_2}) for the same three values of $M_{H^\pm}$: 100 GeV (red wide region), 200 GeV (orange narrower region) and 500 GeV (cyan narrowest region). We can thus observe the expected behaviour, as the charged Higgs contribution becomes smaller for larger masses.

\begin{figure}[!t]
\centering
\includegraphics[scale=0.45]{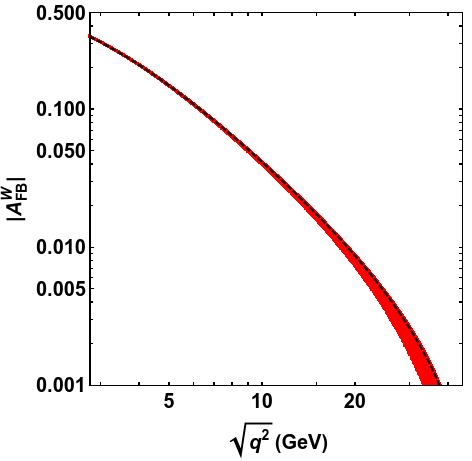}  \quad \includegraphics[scale=0.44]{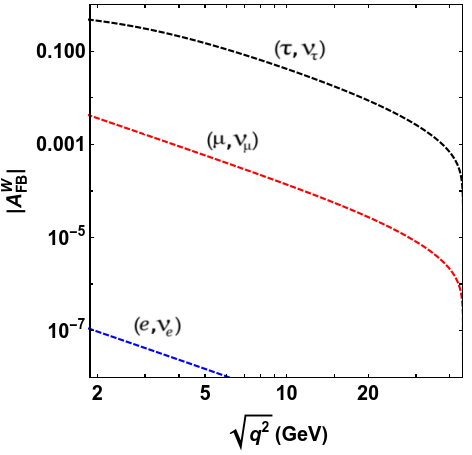}
\caption{Left: $|\mathcal{A}_{\rm{FB}}^{W}|$ as a function of $\sqrt{q^2}$ for $(\tau^+, \, \nu_\tau)$ final state fermions for the SM (dashed curve) and the ATHDM predictions (red region).
Right: SM prediction for $|\mathcal{A}_{\rm{FB}}^{W}|$ for $(\tau^+, \, \nu_\tau)$, $(\mu^+, \, \nu_\mu)$ and $(e^+, \, \nu_e)$ final state fermions.}
\label{AfbW}
\end{figure}

In Fig.~\ref{AfbW} (left) we show the absolute values\footnote{Its values are negative for the whole range of $\sqrt{q^2}$.} of $\mathcal{A}_{\rm{FM}}^{W}$ in double logarithmic scale for the ATHDM for $M_{H^\pm} \in [100,\, 500]$ (GeV), in red. The black-dashed curve represents the SM prediction. It can be observed that the absolute values are rather small. It is also worth mentioning that SM loop corrections (even if they are expected to be small) might also produce non-negligible deviations from the tree-level prediction of the FB asymmetry. This would indeed represent an extra challenge in disentangling new physics effects from the SM prediction in this channel. 

Similar plots to the ones shown in Fig.~\ref{yhlconstrW_a_mu} can be obtained for final state electrons or muons, however with smaller absolute values of the FB asymmetry. They are shown in Fig.~\ref{AfbW} (right). Measuring such values is currently unfeasible at present or even near future colliders and shall not be analysed further.

\subsubsection{$Zff'$ Channel}
\label{zff}

In the CP-conserving limit of the potential and CP-violating Yukawa sector, the only non-vanishing $\mathcal{F}_{jk}$ term (\ref{Fjk}) corresponds to $j=k=3$ (where $\varphi_{k=3} = A$). We thus obtain
\begin{align}
\mathcal{F}_{33} = |\varsigma_l|^2 \, \sin^2\tilde{\alpha} \, ,
\end{align}
and the $(\mathcal{R}_{i3}\mathcal{R}_{k2} - \mathcal{R}_{i2}\mathcal{R}_{k3}) \, \text{Re} (y_{l}^{\varphi_k^0} )$ combination from (\ref{B12}), for $k=3$ (and $i=1$), simplifies to 
\begin{align}
(\mathcal{R}_{13}\mathcal{R}_{32} - \mathcal{R}_{12}\mathcal{R}_{33}) \, \text{Re} (y_{l}^A ) = \text{Im} (\varsigma_l) \, \sin\tilde{\alpha} \, .
\end{align}
Also the $\mathcal{G}_{ij}$ term (\ref{Gij}) reduces to (for $j = 3$, which is the only contribution in the CP-conserving limit of the potential, as mentioned previously) 
\begin{align}
\mathcal{G}_{13}(q^2) = \frac{-\sin\tilde{\alpha}}{q^2 - M_A^2} \, .
\end{align}

\begin{figure}[!t]
\centering
\includegraphics[scale=0.48]{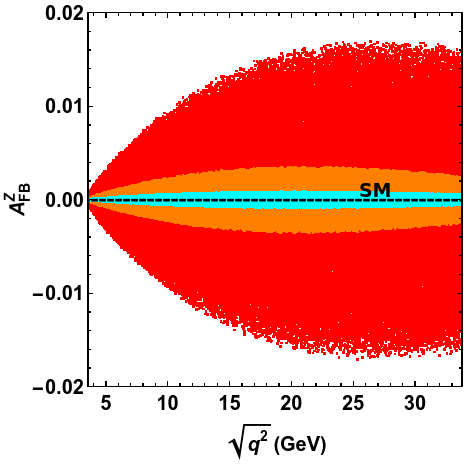}  \qquad \includegraphics[scale=0.48]{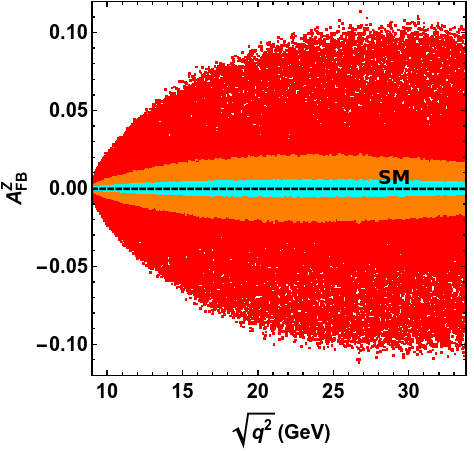}
\caption{Left: $\mathcal{A}_{\rm{FB}}^{Z}$ as a function of $\sqrt{q^2}$ for $(\tau^+, \, \tau^-)$ final state fermions for the ATHDM with $M_A$ 50 GeV (wide red region), 100 GeV (orange intermediate region) and 200 GeV (cyan narrow region). Right: same, for $(b, \, \bar{b})$ final state fermions.}
\label{AfbZ}
\end{figure}

In what follows, let us numerically analyse the ATHDM prediction for the FB asymmetry for $(\tau^+, \, \tau^- )$ and $(b, \, \bar{b})$ final state fermions. In Fig.~\ref{AfbZ} (left) we observe that the prediction for the asymmetry reaches up to 1.6\% for $(\tau^+, \, \tau^- )$ and, up to 10\% for $(b, \, \bar{b} )$ (Fig.~\ref{AfbZ}, right) at high $q^2$, in both cases for a low value of $M_A$ (50 GeV). Unlike the previous case (for $V=W$), the contribution of the $\Lambda_{\text{FB}}^Z(q^2,\cos\theta)$ term is kinematically suppressed and so, the asymmetry rapidly decreases for higher masses of $A$, as it can be observed in the same figure. For the remaining fermions the values of $\mathcal{A}_{\text{FB}}$ are much smaller and are not shown. 

We can thus conclude, as already mentioned previously, that experimentally measuring such an asymmetry in the $Z$ channel, as it is absent in the SM, would offer hints on both extra sources of CP-violation and on the existence of additional scalar bosons.

\section{Conclusions}

In this work we have analysed the FB asymmetry for the three body decay $h\to V ff'$ where $V=W, \,Z$ and, $h$ is the lightest CP-even scalar in the ATHDM. Additionally, we have considered a CP-violating Yukawa sector. In order to obtain the allowed experimental regions for the extra parameters we have performed a fit with the minimal needed experimental data, corresponding to the LHC signal strengths only, as the parameter space was very reduced in this case.   

We have shown that, as expected, the decay rates, and by extension the strengths, are practically unaffected by the (Yukawa suppressed) extra contributions from Fig.~\ref{diagramsh4f}, diagrams $(1')$ and $(2)$. We have seen however, that these diagrams are extremely relevant when studying the FB asymmetry. Particularly, for $V=W$ we have seen that, contrary to one would expect, the contributions from diagram $(1')$ are extremely relevant and {dominate} over $(2)$ in most regions of the parameter space.

Another important prediction is the FB asymmetry in the ATHDM for the $Z$ channel, which is absent in the SM. As this asymmetry is achievable in the presence of extra CP-violating sources (in this case for CP-violation in the Yukawa sector), measuring non-zero values in any region of $q^2$ in this channel, would constitute an unmistakable signature of beyond SM physics.

One should also note the following. The same way the presence of extra scalars can affect the decay $\varphi_i^0 \to Vff'$ is can also affect the production of $\varphi_i^0$, i.e, through the vector-boson fusion cross section or via Higgs-strahlung. However, as we have seen, the total decay widths are hardly affected by the presence of extra scalars. The only sizeable effect appears for the  the FB asymmetry, whose contribution to the total decay rate vanishes when integrating over the polar $\theta$ angle. Even if these are not strict arguments, and in order to assure such affirmations one should perform the complete calculation, in the presence of the previously obtained results, one can be rather confident that the current approximations of the cross sections, that do not include the extra contributions in the THDM, are valid.

Last, a brief summary on the interpretation of the previously obtained equations in terms of a CP-violating potential (with small CP-violating effects) and real Yukawa couplings can be found in Appendix~\ref{CP-pot}. If a deviation (from the SM) of the FB asymmetry is finally measured, reinterpreting the results in this scenario would translate into additional constrains on the masses of the scalar bosons and on the scalar potential parameters. Of course one can consider both a CP-violating scalar potential and Yukawa sector, however such an analysis is far beyond the goal of this paper.

\begin{appendix}
\section{Three-body kinematics}
\label{AppA}

\begin{figure}[!htb]
\centering
\includegraphics[scale=0.5]{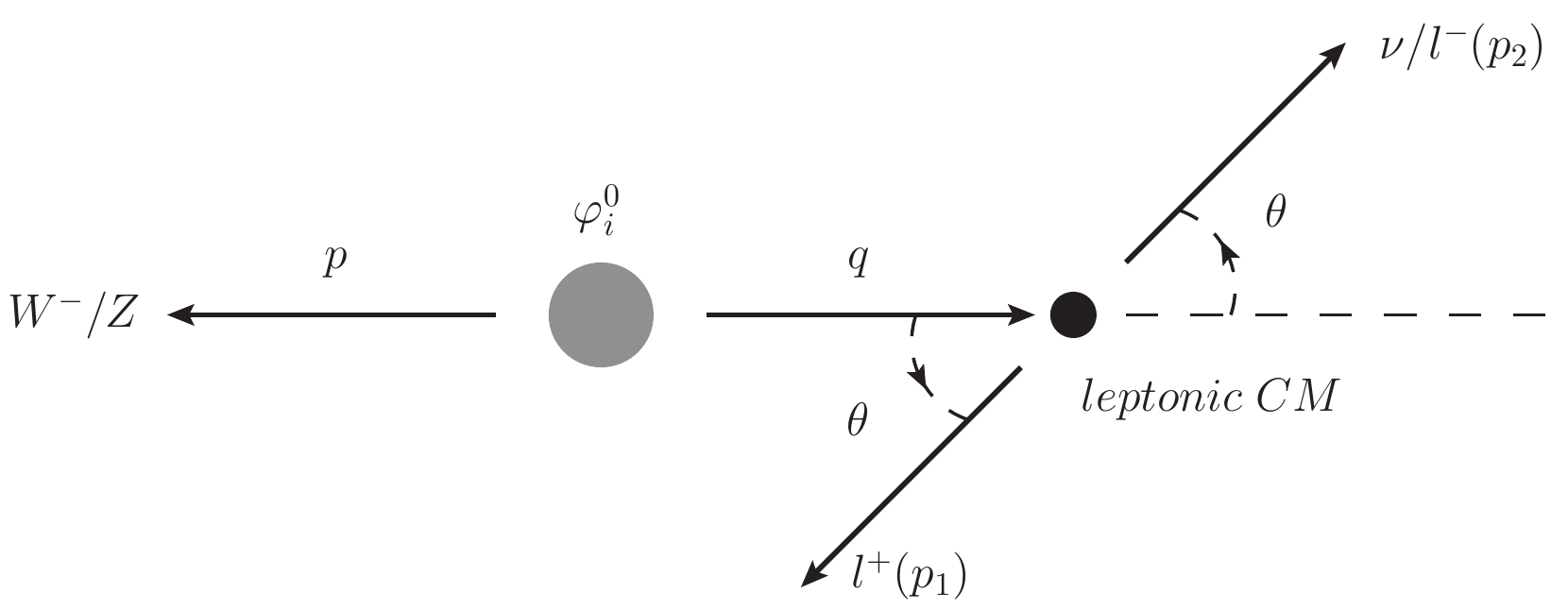}
\caption{Momenta assignation and the polar $\theta$ angle defined in the leptonic CM.}
\label{momenta}
\end{figure}

The momenta assignation and the polar $\theta$ angle defined in the leptonic CM are shown in Fig.~\ref{momenta}. The experimentally measurable angular distribution thus corresponds to the momentum transfer distribution $(p_{\varphi_i^0} - p_{l^+})^2$, or equivalently the positive lepton energy distribution in the Higgs rest reference frame.

For the squared amplitude calculation, the following results were used
\begin{align}
&& p\cdot q &= \frac{1}{2} (M_{\varphi_i^0}^2 - M_{W/Z}^2 - q^2 ) \,,  &&& p_1\cdot q &= \frac{1}{2} (q^2 + m_l^2 - m_{\nu/l}^2 ) \, , && \notag  \\
&& p_2\cdot q &= \frac{1}{2} ( q^2 - m_l^2 + m_{\nu/l}^2  ) \,,  &&& p_1\cdot p_2 &= \frac{1}{2} (q^2 - m_l^2 - m_{\nu/l}^2 ) \, , &&
\end{align}
and
\begin{align}
p\cdot p_1 = p\cdot q - p\cdot p_2 \, , 
\end{align}
where the scalar product $p\cdot p_2$ is given by 
\begin{align}
p\cdot p_2 &= \frac{1}{4 q^2} ( M_{\varphi_i^0}^2-M_W^2-q^2 ) ( q^2 - m_l^2 + m_{\nu/l}^2 )  \notag \\
& \qquad + \frac{1}{4q^2}\lambda^{1/2} (M_{\varphi_i^0}^2,M_W^2,q^2) \lambda^{1/2} (q^2,m_l^2,m_{\nu/l}^2) \cos \theta \, .
\label{scal_prod}
\end{align}
with $m_\nu = 0$ for the neutrino mass. Finally, the thee-body phase space, given by 
\begin{align}
dQ_3 = \frac{1}{(2\pi)^5} \frac{d^3 p}{2p^0}\frac{d^3 p_1}{2p_1^0}\frac{d^3 p_2}{2p^0_2} \delta^{(4)}(k-p-p_1-p_2) \, ,
\end{align}
(where $k^\mu$ is the scalar boson four-momentum) takes the following simple form when expressed in terms of $q^2$ and $\cos\theta$
\begin{align}
dQ_3 = \frac{1}{256 \, \pi^3 M_{\varphi_i^0}^2 \, q^2} \lambda^{1/2} (M_{\varphi_i^0}^2,M_W^2,q^2) \lambda^{1/2}(q^2,m_l^2,m_{\nu/l}^2) \, dq^2 \, d\cos\theta \, .
\end{align}

\section{Three-body differential decay rates}
\label{AppForm}

In order to insure the correctness of the following analytical results, as a cross check, the calculations have been first performed, partially on paper and partially with Feyncalc (traces, simplifications with Mathematica, etc.,), and second, with a fully automatized Feyncalc routine. The obtained results were of course identical. As a second cross check, when integrating over $q^2$ and $\cos\theta$, one obtains in the $m_f \to 0$ and $\mathcal{R}_{i1} \to 1$ limit the SM total three-body decay width \cite{Keung,Djouadi:2005gi} for both $W$ and $Z$ channels. Additionally, the rather complicated structure of $\Lambda_{\text{FB}}(q^2)$ corresponding to the $Z$ channel vanishes when integrated, which is needed in order to obtain a null contribution to the $\mathcal{A}_{\text{FB}}$ asymmetry in the SM limit.

\subsection{$\mathbf{\Gamma(\varphi_i^0 \to W^- l^+\nu)}$}

The angular coefficients, corresponding to the contributions of diagrams $(1)$ and $(2)$ to the double differential decay rate (\ref{diffDecay_2}) for $V=W^-$, can be expressed as sums of the the terms corresponding to each of the two squared amplitudes (indices 1 and 2) and the crossed term (indices 12):
\begin{gather}
a^W(q^2) = \mathcal{C}^W  \left[ \mathcal{A}_1^W \, a_1^W  +  \mathcal{A}_2^W \, a_2^W + \mathcal{A}_{12}^W \, a_{12}^W  \right] \, , \qquad 
b^W(q^2) = \mathcal{C}^W \left[ \mathcal{A}_1^W \, b_1^W +   \mathcal{A}_{12}^W \, {b}_{12}^W \right] \, ,
 \notag \\[1.5ex]
c^W(q^2) = \mathcal{C}^W  \mathcal{A}_1^W \, c_1^W  \, .
\end{gather}
The common factor $\mathcal{C}^W$ from the previous expressions is originated from the three-body phase-space (see Appendix~\ref{AppA}) times the $1/(2M_{\varphi_i^0})$ factor corresponding to the definition of the decay rate. Its explicit expression reads
\begin{align}
\label{CW}
\mathcal{C}^W &= \frac{1}{512 \pi^3 M_{\varphi_i^0}^3 \, q^2} \lambda^{1/2} (M_{\varphi_i^0}^2, M_W^2, q^2) \, \lambda^{1/2} (q^2,m_l^2, 0)   \, ,
\end{align}
where $\lambda(x,y,z)\equiv x^2+y^2+z^2 - 2xy - 2xz -2yz$ is the usual Kallen lambda function. The terms corresponding to the squared amplitude of the first diagram are given by
\begin{align}
\mathcal{A}_1^W = \mathcal{R}_{i1}^2 \frac{g^2}{2} \frac{M_W^2}{v^2} \frac{1}{(q^2-M_W^2)^2} \, , 
\end{align}
and
\begin{align}
{a}_1^W &= q^2(q^2 + 6M_W^2-2M_{\varphi_i^0}^2) + (M_W^2 - M_{\varphi_i^0}^2)^2 \left( 1 -\frac{m_l^4}{q^4} \right)  \, ,
\notag \\
b_1^W &= \frac{2 m_l^2 ( M_W^2 - q^2 )}{q^4 \, M_W^2}  \,
(q^2 + M_W^2 - M_{\varphi_i^0}^2 ) \, \lambda^{1/2} (M_{\varphi_i^0}^2, M_W^2, q^2) \, \lambda^{1/2} (q^2,m_l^2, 0)  \, ,  \notag \\
\label{abc1w}
{c}_1^W &= -\frac{1}{q^4} \lambda (M_{\varphi_i^0}^2, M_W^2, q^2) \, \lambda (q^2,m_l^2, 0) \, .
\end{align}
For the second diagram, for the corresponding squared amplitude, we only find terms that contribute to $a^W$
\begin{align}
\mathcal{A}_2^W &=  |\varsigma_l|^2 \, (\mathcal{R}_{i2}^2 + \mathcal{R}_{i3}^2) \, \frac{m_l^2}{v^2} \, \frac{g^2}{2}  \frac{1}{(q^2-M_{H^\pm}^2)^2} \, , \notag \\
{a}_2^W &= \frac{(q^2-m_l^2)}{M_W^2} \left[ (M_W - M_{\varphi_i^0})^2 - q^2 \right] \left[ (M_W + M_{\varphi_i^0})^2 - q^2 \right]  \, .
\label{abc2w}
\end{align}
which is due to the fact that a scalar boson decaying into a pair of fermions does not generate any angular dependence. Finally, the crossed term gives the following contributions to $a^W$ and $b^W$
\begin{align}
\mathcal{A}_{12}^W &=   \frac{m_l^2}{v^2} \, g^2 \, \text{Re} \Big[ \varsigma_l^* \mathcal{R}_{i1} (\mathcal{R}_{i2} - i \mathcal{R}_{i3} ) \Big] \, \frac{1}{(q^2-M_W^2)(q^2-M_{H^\pm}^2)} \, , \notag  \\
{a}_{12}^W &= \frac{(M_W^2 - q^2)(m_l^2-q^2)}{q^2 \, M_W^2} \left[ (M_W - M_{\varphi_i^0})^2 - q^2 \right] \left[ (M_W + M_{\varphi_i^0})^2 - q^2 \right]    \, , \notag  \\
b_{12}^W &= -\frac{(q^2+M_W^2-M_{\varphi_i^0}^2)}{q^2} \, \lambda^{1/2} (M_{\varphi_i^0}^2, M_W^2, q^2) \, \lambda^{1/2} (q^2, m_l^2, 0)  \, .
\label{abc12w}
\end{align}

Let us now calculate the remaining terms, that will contribute to $\Lambda^W(q^2,\cos\theta)$. We obtain the following structure
\begin{align}
a^W_\Lambda(q^2,\cos\theta) &= \mathcal{C}^W  \left[ \mathcal{A}_{1'}^W \, a_{1'}^W  +  \mathcal{A}_{1'1}^W \, a_{1'1}^W  +  \mathcal{A}_{1'2}^W \, a_{1'2}^W  \right] \, , \notag \\[1.5ex] 
b^W_\Lambda(q^2,\cos\theta) &= \mathcal{C}^W  \left[ \mathcal{A}_{1'}^W \, b_{1'}^W  +  \mathcal{A}_{1'1}^W \, b_{1'1}^W  +  \mathcal{A}_{1'2}^W \, b_{1'2}^W  \right] \, , \notag \\[1.5ex]
c^W_\Lambda(q^2,\cos\theta) &= \mathcal{C}^W  \left[ \mathcal{A}_{1'}^W \, c_{1'}^W  +  \mathcal{A}_{1'1}^W \, c_{1'1}^W     \right]  \,  .
\end{align}
Corresponding to the squared amplitude of diagram $(1')$ we obtain 
\begin{align}
\mathcal{A}_{1'}^W = \frac{g^2}{8} \frac{m_l^2}{v^2} \frac{1}{(M_W^2 - m_l^2 + 2\, p \cdot p_2)^2} \Big[ \text{Re} \Big( y_{l}^{\varphi_i^0} \Big)^2 + \text{Im} \Big( y_{l}^{\varphi_i^0} \Big)^2  \Big] \, , 
\label{prop_prime}
\end{align}
where the scalar product $p\cdot p_2$ is given by (\ref{scal_prod}) with $m_\nu = 0$
\begin{align}
p\cdot p_2 &= \frac{1}{4 q^2} ( M_{\varphi_i^0}^2-M_W^2-q^2 ) ( q^2 - m_l^2)  \notag \\ & 
\qquad \qquad \qquad + \frac{1}{4q^2}\lambda^{1/2} (M_{\varphi_i^0}^2,M_W^2,q^2) \lambda^{1/2} (q^2,m_l^2,0) \cos \theta \, .
\label{scal_prod_}
\end{align}
The rest of the terms corresponding to this diagram read
\begin{align}
{a}_{1'}^W &= \frac{1}{M_W^2} \left[ q^2 (q^4 - 2 q^2 M_{\varphi_i^0}^2 -  2 M_W^2 M_{\varphi_i^0}^2 - 3 M_W^4  + M_{\varphi_i^0}^4 ) + 2(M_W^3 - M_W M_{\varphi_i^0}^2)^2 \right]  
\notag \\
& \qquad\qquad  - \frac{2 m_l^4}{q^4} (M_W^2-M_{\varphi_i^0}^2)^2  \, , \notag \\
b_{1'}^W &=  \left[ \frac{2(q^2 - M_W^2 - M_{\varphi_i^0}^2)}{M_W^2}  + \frac{4m_l^2 (M_W^2 - M_{\varphi_i^0}^2 )}{q^4}\right] \, \lambda^{1/2} (M_{\varphi_i^0}^2, M_W^2, q^2) \, \lambda^{1/2} (q^2,m_l^2, 0)  \, ,  \notag \\
\label{abc1w}
{c}_{1'}^W &= \frac{q^2 - 2 M_W^2}{M_W^2 \, q^4} \, \lambda (M_{\varphi_i^0}^2, M_W^2, q^2) \, \lambda (q^2,m_l^2, 0) \, .
\end{align}
The expressions corresponding to the crossed term of diagrams $(1)$ and $(1')$ are given by ($q\equiv \sqrt{q^2}$)
\begin{align}
\mathcal{A}_{1'1}^W &= -\frac{g^2}{2} \frac{m_l^2}{v^2} \mathcal{R}_{i1} \, \text{Re} \Big( y_{l}^{\varphi_i^0} \Big) \frac{1}{(q^2 - M_W^2)(M_W^2 - m_l^2 + 2\, p \cdot p_2)}  \, , \notag \\
        {a}_{1'1}^W &= \frac{(m_l^2 - q^2)}{M_W^2 \, q^4} \bigg[ 2m_l^2 \Big( M_W^4 (q^2 - 2 M_{\varphi_i^0}^2) + M_W^2 (M_{\varphi_i^0}^4 - 3q^4) + M_W^6 - ({q}^3 - M_{\varphi_i^0}^2 {q}  )^2  \Big) \notag \\
        & \qquad\qquad + q^2   \Big(  M_W^4 (11 q^2 - 6 M_{\varphi_i^0}^2) - 2M_W^2 q^2  (M_{\varphi_i^0}^2 + q^2) + 6 M_W^6 +     ({q}^3 - M_{\varphi_i^0}^2 {q}  )^2         \Big)   \bigg] \, , \notag \\
        {b}_{1'1}^W &= -\lambda^{1/2}(M_{\varphi_i^0}^2,M_W^2, q^2) \lambda^{1/2}(q^2,m_l^2,0) \bigg\{ \frac{4m_l^2}{q^4} (M_W^2 - M_{\varphi_i^0}^2) \notag \\ 
        & \qquad \qquad \qquad \qquad   +\frac{1}{M_W^2 \, q^2} \Big[ q^4 + 2M_W^2 (2M_W^2 + M_{\varphi_i^0}^2) - q^2 (3M_W^2 + M_{\varphi_i^0}^2)   \Big]  \bigg\} \, , \notag \\
        {c}_{1'1}^W &= \frac{2}{q^4} \lambda(M_{\varphi_i^0}^2,M_W^2, q^2) \lambda(q^2,m_l^2,0) \, .
\end{align}
Finally, the last contribution, of the remaining crossed term (of diagrams $(2)$ and $(1')$) reads
\begin{align}
\mathcal{A}_{1'2}^W &= \frac{g^2}{2}   \frac{m_l^2}{v^2} \text{Re} \Big[ \varsigma_l^* (\mathcal{R}_{i2} - i \mathcal{R}_{i3} ) \Big] \text{Re} \Big( y_{l}^{\varphi_i^0} \Big)   \frac{1}{(q^2 - M_{H^\pm}^2)(M_W^2 - m_l^2 + 2\, p \cdot p_2)}  \, , \notag \\
        {a}_{1'2}^W &= \frac{(q^4 + 2m_l^4 - 3m_l^2 q^2)}{q^2 \, M_W^2} \left[ (M_W - M_{\varphi_i^0})^2 - q^2 \right] \left[ (M_W + M_{\varphi_i^0})^2 - q^2 \right] \, , \\
        {b}_{1'2}^W &= \left[ \frac{(q^2 - M_W^2 - M_{\varphi_i^0}^2)}{M_W^2} - \frac{2m_l^2 (M_W^2 - M_{\varphi_i^0}^2)}{q^2 \, M_W^2} \right] \lambda^{1/2}(M_{\varphi_i^0}^2,M_W^2, q^2) \lambda^{1/2}(q^2,m_l^2,0)  \, . \notag 
\end{align}

In the following we shall analyse the formulae of the differential decay rate corresponding to the ${\varphi_i^0 \to Zl^+l^-}$ process.

\subsection{$\mathbf{\Gamma(\varphi_i^0 \to Zl^+l^-)}$}

Similar to the previous case, we write the angular coefficients for the $V=Z$ case as
\begin{align}
a^Z(q^2) = \mathcal{C}^Z \, \left[ \mathcal{A}_1^Z \, a_1^Z +  \mathcal{A}_2^Z \, a_2^Z  \right] \, , \qquad  b^Z(q^2) = \mathcal{C}^Z \, \mathcal{B}_{12}^Z   \, , \qquad
c^Z(q^2) = \mathcal{C}^Z \, \mathcal{A}_1^Z \, c_1^Z  \, .
\end{align}
The common factor $\mathcal{C}^Z$ explicitly reads
\begin{align}
\mathcal{C}^Z = \frac{1}{512 \pi^3 M_{\varphi_i^0}^3} \lambda^{1/2} (M_{\varphi_i^0}^2, M_Z^2, q^2)  \lambda^{1/2} (q^2,m_l^2, m_l^2)\, , 
\end{align}
and, the remaining terms corresponding to the squared amplitude of the first diagram are given by
\begin{align}
\mathcal{A}_1^Z &= \mathcal{R}_{i1}^2 \frac{g^2}{2c_{\text{w}}^2} \frac{M_Z^2}{v^2}  \frac{(a_l^2 + v_l^2)}{(q^2-M_Z^2)^2} \, , \notag  \\
{a}_1^Z &= q^2(q^2 + 6M_Z^2-2M_{\varphi_i^0}^2) + (M_Z^2 - M_{\varphi_i^0}^2)^2 \, , \\
{c}_1^Z &=  -\frac{1}{q^4}\lambda (M_{\varphi_i^0}^2, M_Z^2, q^2) \lambda (q^2, m_l^2, m_l^2) \, . \notag
\end{align}
Note that in this case we have no contribution from the first diagram to the $b^Z$ term. As for the second diagram, for the same reason as in the previous case (for $V=W$) we only obtain a term contributing to $a^Z$
\begin{align}
\mathcal{A}_2^Z &= \sum_{j,k}  \frac{g^2}{2c_{\text{w}}^2}\frac{m_l^2}{v^2} \frac{\mathcal{F}_{jk} (q^2)}{(q^2-M_{\varphi_j^0}^2)(q^2-M_{\varphi_k^0}^2)} \, , \notag \\
{a}_2^Z &= \frac{ 1 }{M_Z^2} \left[ (M_Z - M_{\varphi_i^0})^2 - q^2 \right] \left[ (M_Z+ M_{\varphi_i^0})^2 - q^2 \right] \, , 
\end{align}
where the explicit expression for $\mathcal{F}_{jk}(q^2)$ reads
\begin{align}
\mathcal{F}_{jk} (q^2) &= (\mathcal{R}_{i3}\mathcal{R}_{j2} - \mathcal{R}_{i2}\mathcal{R}_{j3})(\mathcal{R}_{i3}\mathcal{R}_{k2} - \mathcal{R}_{i2}\mathcal{R}_{k3}) 
\left( q^2 (\mathcal{Y}_{jk}^{\text{R}} + \mathcal{Y}_{jk}^{\text{I}}) - 4 m_l^2 \, \mathcal{Y}_{jk}^{\text{R}} \right) \, ,
 \label{Fjk}
\end{align}
with $\mathcal{Y}_{jk}^{\text{R}}$ and $\mathcal{Y}_{jk}^{\text{I}}$ encoding the following combinations of Yukawa couplings
\begin{align}
\mathcal{Y}_{jk}^{\text{I}} = \text{Im} \Big( y_{l}^{\varphi_j^0} \Big) \text{Im} \left( y_{l}^{\varphi_k^0} \right) \, , \qquad
\mathcal{Y}_{jk}^{\text{R}} = \text{Re} \Big( y_{l}^{\varphi_j^0} \Big) \text{Re} \left( y_{l}^{\varphi_k^0} \right) \, .
\label{YUK}
\end{align} 
Finally, the contribution of the crossed term reads
\begin{align}
\mathcal{B}_{12}^Z &= 2 \, v_l \, \mathcal{R}_{i1} \frac{g^2}{c_{\text{w}}^2} \, \frac{m_l^2}{v^2} \,  \frac{(q^2 + M_Z^2 - M_{\varphi_i^0}^2)}{q^2(q^2-M_Z^2)} \lambda^{1/2}(M_{\varphi_i^0}^2, M_Z^2, q^2) \lambda^{1/2}(q^2, m_l^2, m_l^2)  \notag \\ 
& \qquad \qquad \qquad \qquad \qquad \qquad \times \sum_k  \frac{1}{q^2 - M_{\varphi_k^0}^2} (\mathcal{R}_{i3}\mathcal{R}_{k2} - \mathcal{R}_{i2}\mathcal{R}_{k3}) \, \text{Re} \left( y_{l}^{\varphi_k^0} \right) \, .
\label{B12}
\end{align}
The neutral current couplings $v_l$ and $a_l$ are, as usual given by ($s_{\text{w}} \equiv \sin\theta_{\text{w}}$, $c_{\text{w}} \equiv \cos\theta_{\text{w}}$ with $\theta_{\text{w}}$ the weak mixing angle)
\begin{align}
v_l = -\frac{1}{2} + 2 \, s_{\text{w}}^2 \, , \qquad \qquad a_l = -\frac{1}{2} \, .
\end{align}

The terms that involve the two diagrams for $V=Z$ shown $(1')$, that contribute to the $\Lambda^Z(q^2,\cos\theta)$ function, can be grouped as
\begin{align}
a^Z_\Lambda(q^2,\cos\theta) &= \mathcal{C}^Z  \left[ \mathcal{A}_{1'}^Z \; (a_{1'}^Z + \bar a_{1'}^Z )  +  \mathcal{A}_{1'1}^Z \, a_{1'1}^Z  +  \mathcal{A}_{1'2}^Z \, a_{1'2}^Z  \right] \, , \notag \\[1.5ex] 
b^Z_\Lambda(q^2,\cos\theta) &= \mathcal{C}^Z  \left[ \mathcal{B}_{1'}^Z \; b_{1'}^Z  +  \mathcal{A}_{1'1}^Z \, b_{1'1}^Z  +  \mathcal{A}_{1'2}^Z \, b_{1'2}^Z  \right] \, , \notag \\[1.5ex]
c^Z_\Lambda(q^2,\cos\theta) &= \mathcal{C}^Z  \left[ \mathcal{C}_{1'}^Z \; (c_{1'}^Z + \bar c_{1'}^Z)  +  \mathcal{A}_{1'1}^Z \, c_{1'1}^Z     \right]  \,  .
\label{LambdaFB}
\end{align}
The squared amplitude of diagram $(1')$ brings the following contribution
\begin{align}
\mathcal{A}_{1'}^Z &= \frac{g^2}{4 \, c_{\text{w}}^2} \frac{m_l^2}{v^2} \, (\mathcal{Y}^\text{R}_{ii} + \mathcal{Y}^\text{I}_{ii}) \, \frac{1}{M_Z^2}  \, , \notag \\
a_{1'}^Z & = \frac{1}{2}\left[ \frac{(a_l^2 + v_l^2)}{(M_{\varphi_i^0}^2 - q^2 - 2p\cdot p_2)^2} + \frac{(a_l^2 + v_l^2)}{(M_Z^2 + 2p\cdot p_2)^2}\right] \notag \\
&  \qquad \times \left[-M_Z^4 (4 M_{\varphi_i^0}^2 + 3 q^2) + 2 M_Z^2 (M_{\varphi_i^0}^4 - M_{\varphi_i^0}^2  \, q^2) + 2 M_Z^6 + (q^3 - M_{\varphi_i^0}^2 q)^2 \right]
 \\
\bar{a}_{1'}^Z & = -\frac{(a_l^2 - v_l^2)}{(M_{\varphi_i^0}^2 - q^2 - 2p\cdot p_2)(M_Z^2 + 2p\cdot p_2)} \notag 
\\
& \qquad  \times \left[ -M_Z^4 (4 M_{\varphi_i^0}^2 + 5 q^2)+2 M_Z^2 (M_{\varphi_i^0}^2 \, q^2 + M_{\varphi_i^0}^4 + 2 q^4)+2 M_Z^6 - (q^3 - M_{\varphi_i^0}^2 q)^2 \right] \, , \notag 
\end{align}
with the scalar product $p\cdot p_2$ again, given by (\ref{scal_prod}) which in this case reads
\begin{align}
p\cdot p_2 &= \frac{1}{4} ( M_{\varphi_i^0}^2-M_Z^2-q^2 ) + \frac{1}{4q^2}\lambda^{1/2} (M_{\varphi_i^0}^2,M_Z^2,q^2) \lambda^{1/2} (q^2,m_l^2,m_l^2) \cos \theta \, ,
\end{align}
and where $\mathcal{Y}^\text{I}_{ii}$ and $\mathcal{Y}^\text{R}_{ii}$ are given in (\ref{YUK}) for $j=i=k$. The remaining terms of the squared elements of $(1')$ read
\begin{align}
\mathcal{B}_{1'}^Z &= \frac{g^2}{4 \, c_{\text{w}}^2} \frac{m_l^2}{v^2} \,  \frac{1}{M_Z^2 \, q^2}
\lambda^{1/2}(M_{\varphi_i^0}^2, M_Z^2, q^2)   \lambda^{1/2}(q^2, m_l^2, m_l^2)  \, , \notag \\
b_{1'}^Z &= \left[ \frac{1}{(M_{\varphi_i^0}^2 - q^2 - 2p\cdot p_2)^2} - \frac{1}{(M_Z^2 + 2p\cdot p_2)^2}\right] \notag \\ 
&  \qquad \qquad \times \bigg\{
q^2(a_l^2 + v_l^2) \left[ (\mathcal{Y}^\text{R}_{ii} + \mathcal{Y}^\text{I}_{ii}) (M_Z^2 + M_{\varphi_i^0}^2 - q^2) + 4m_l^2 \mathcal{Y}^\text{R}_{ii} \right]
\notag \\
& \qquad\qquad \qquad\qquad \qquad\qquad - 4 m_l^2 \mathcal{Y}^\text{R}_{ii} \left[  a_l^2 (5M_Z^2 + M_{\varphi_i^0}^2) + v_l^2 (M_{\varphi_i^0}^2 -M_Z^2)  \right]
 \bigg\} \, , 
\label{b1p}
\end{align}
and finally
\begin{align}
\mathcal{C}_{1'}^Z &= \frac{g^2}{4 \, c_{\text{w}}^2} \frac{m_l^2}{v^2} \, \frac{1}{M_Z^2 \, q^4} \, \lambda(M_{\varphi_i^0}^2, M_Z^2, q^2)  \lambda(q^2, m_l^2, m_l^2)  \, , \notag \\ 
{c}_{1'}^Z &=- \frac{1}{2}\left[ \frac{(a_l^2 + v_l^2)}{(M_{\varphi_i^0}^2 - q^2 - 2p\cdot p_2)^2} + \frac{(a_l^2 + v_l^2)}{(M_Z^2 + 2p\cdot p_2)^2}\right] \bigg[
(\mathcal{Y}^\text{R}_{ii} + \mathcal{Y}^\text{I}_{ii}) (2 M_Z^2 - q^2) + 4m_l^2 \mathcal{Y}^\text{R}_{ii}  \bigg] \, , 
\notag \\
\bar{c}_{1'}^Z &= \frac{1}{(M_{\varphi_i^0}^2 - q^2 - 2p\cdot p_2)(M_Z^2 + 2p\cdot p_2)} \notag \\ &  \qquad \qquad\qquad \qquad \times \bigg[ (a_l^2 - v_l^2)
(\mathcal{Y}^\text{R}_{ii} + \mathcal{Y}^\text{I}_{ii}) (2 M_Z^2 - q^2) - 4m_l^2 (a_l^2 + v_l^2)\mathcal{Y}^\text{R}_{ii}  \bigg] \, .
\end{align}
For the crossed term corresponding to diagrams $(1)$ and $(1')$ we obtain
\begin{align}
\mathcal{A}_{1'1}^Z &= -\frac{g^2}{c_{\text{w}}^2} \mathcal{R}_{i1} \, \text{Re} \Big( y_{l}^{\varphi_i^0} \Big) \, \frac{m_l^2}{v^2} \frac{1}{(q^2 - M_Z^2)} \left[ \frac{1}{(M_{\varphi_i^0}^2 - q^2 - 2p\cdot p_2)} + \frac{\epsilon_{1'1}^J}{(M_Z^2 + 2p\cdot p_2)}\right] \,  , \notag \\
{a}_{1'1}^Z & = \frac{1}{M_Z^2}  \bigg\{ a_l^2 (4 m_l^2 - q^2) \left[ -2 M_Z^2 ( M_{\varphi_i^0}^2 + q^2) + 
     9 M_Z^4 + ( M_{\varphi_i^0}^2 - q^2)^2 \right] \notag \\
     & \qquad\qquad\qquad\qquad\qquad\qquad\qquad\qquad\qquad - 
  2 v_l^2 \, M_Z^4  (3 M_Z^2 - 3  M_{\varphi_i^0}^2 + q^2) \bigg\} \, , \notag \\
{b}_{1'1}^Z & = \frac{-1}{q^2 \, M_Z^2} \lambda^{1/2}(M_{\varphi_i^0}^2, M_Z^2, q^2)   \lambda^{1/2}(q^2, m_l^2, m_l^2) 
\bigg\{ v_l^2 \,  M_Z^2( M_Z^2 - M_{\varphi_i^0}^2 + q^2 )  \notag \\ & \qquad\qquad\qquad\qquad\qquad\qquad - a_l^2 \left[M_Z^2 ( M_{\varphi_i^0}^2 - 2 q^2 ) + 5 M_Z^4 - M_{\varphi_i^0}^2 q^2 + q^4 \right]  \bigg\}  \, , \notag \\
{c}_{1'1}^Z & = \frac{1}{q^4} (a_l^2 + v_l^2) \lambda(M_{\varphi_i^0}^2, M_Z^2, q^2)   \lambda(q^2, m_l^2, m_l^2) \, ,
\label{b11p}
\end{align}
where $\epsilon_{1'1}^J = -1$ for $J={b}_{1'1}^Z$, and $+1$ for the rest. The last contribution corresponds to the crossed term of diagrams $(2)$ and $(1')$. It reads 
\begin{align}
\mathcal{A}_{1'2}^Z &= \frac{g^2}{2c_{\text{w}}^2} \frac{v_l}{M_Z^2} \frac{m_l^2}{v^2}\left[ \frac{1}{(M_{\varphi_i^0}^2 - q^2 - 2p\cdot p_2)} + \frac{\epsilon_{1'2}^J}{(M_Z^2 + 2p\cdot p_2)}\right] \,  , \notag \\
{a}_{1'2}^Z & =  \sum_j \mathcal{G}_{ij}(q^2) \left[ (M_W - M_{\varphi_i^0})^2 - q^2 \right] \left[ (M_W + M_{\varphi_i^0})^2 - q^2 \right] \left[ \mathcal{Y}_{ij}^\text{I} \, q^2 + \mathcal{Y}_{ij}^\text{R} (q^2 -4m_l^2) \right] \, , \notag \\
{b}_{1'2}^Z & =  \sum_j \mathcal{G}_{ij}(q^2)  \left[  
(\mathcal{Y}_{ij}^\text{I} + \mathcal{Y}_{ij}^\text{R}) (M_Z^2 + M_{\varphi_i^0}^2 - q^2) + \frac{4m_l^2}{q^2} \mathcal{Y}_{ij}^\text{R} (M_Z^2 - M_{\varphi_i^0}^2 + q^2 ) 
 \right] \notag \\
 & \qquad\qquad\qquad\qquad\qquad\qquad\qquad\qquad \times \lambda^{1/2}(M_{\varphi_i^0}^2, M_Z^2, q^2)   \lambda^{1/2}(q^2, m_l^2, m_l^2) \, ,
\label{a1p2}
\end{align}
with $\epsilon_{1'2}^J = -1$ for $J={a}_{1'2}^Z$, and $+1$ for the rest and where $\mathcal{G}_{ij}(q^2)$ is given by
\begin{align}
\mathcal{G}_{ij}(q^2) = \frac{( \mathcal{R}_{i3} \mathcal{R}_{j2} - \mathcal{R}_{i2} \mathcal{R}_{j3} )}{q^2 - M_{\varphi_j^0}^2} \, .
\label{Gij}
\end{align}

Let us analyse the contributing terms from (\ref{LambdaFB}) to $\Lambda_{\text{FB}}(q^2)$ defined in (\ref{lamb_fb}). One can check that\footnote{See appendix~\ref{AppInt} for further details.}
\begin{align}
0 &= \left(\int_{-1}^0 - \int_0^1 \right)d\cos\theta \left( \frac{(\cos^2\theta)^m}{(M_{\varphi_i^0}^2 - q^2 - 2p\cdot p_2)^{m'}} + \frac{(\cos^2\theta)^m}{(M_Z^2 + 2p\cdot p_2)^{m'}} \right) \notag \\
  &= \left(\int_{-1}^0 - \int_0^1 \right)d\cos\theta \left( \frac{(\cos^2\theta)^m}{(M_{\varphi_i^0}^2 - q^2 - 2p\cdot p_2)(M_Z^2 + 2p\cdot p_2)} \right) \notag \\
  &= \left(\int_{-1}^0 - \int_0^1 \right)d\cos\theta \left( \frac{\cos\theta}{(M_{\varphi_i^0}^2 - q^2 - 2p\cdot p_2)^{m'}} - \frac{\cos\theta}{(M_Z^2 + 2p\cdot p_2)^{m'}} \right) \, ,
\label{INT}
\end{align}
for any combination of $m$ and $m'$ with $m = 0,1$ and $m'=1,2$. Therefore, the only surviving terms that contribute to $\Lambda_{\text{FB}}(q^2)$ are, as already mentioned, $\mathcal{C}^Z   \mathcal{A}_{1'2}^Z ( a_{1'2}^Z - b_{1'2}^Z \, \cos\theta)$ and so,
\begin{align}
\Lambda_{\text{FB}}(q^2) = \left(\int_{-1}^0 - \int_0^1 \right) \mathcal{C}^Z   \mathcal{A}_{1'2}^Z \, ( a_{1'2}^Z - b_{1'2}^Z \, \cos\theta) \, d\cos\theta \, ,
\end{align}
which is absent in the SM, as it contains contributions from extra scalars. Also as explained in (\ref{Yuk_comb}), in order to have non-vanishing contributions we need extra sources of CP-violation, which in our case translates into complex values of the $\varsigma_{d,l}$ parameters.

For final state (bottom) quarks one can trivially adapt the previous expressions. The needed substitutions are simply
\begin{align}
m_l \to m_b \, , \qquad v_l \to v_d \, , \qquad a_l \to a_d \, , \qquad \varsigma_l \to \varsigma_d \, ,  \qquad y^{\varphi_i^0}_l \to y^{\varphi_i^0}_d  \, ,
\end{align}  
where $v_d$ and $a_d$ are given by
\begin{align}
v_d = -\frac{1}{2} + \frac{2}{3} s^2_{\text{w}} \, , \qquad\qquad a_d = -\frac{1}{2} \, .
\end{align}

\section{LHC signal strengths}
\label{sign}

By integrating the previous differential decay widths one obtains the total decay rates, that are useful in our analysis. When comparing the THDM predictions to the LHC experimental data, one normally re-scales the $hVV$ coupling corresponding to the SM with the factor $R_{i1}$. This means that one implicitly assumes negligible contributions from $H^\pm$ or extra neutral scalars. The more generic formulae that include the additional scalar boson contributions can be trivially deuced, and will be done in the following.

The experimental results for both channels are given in terms of the Higgs production cross section times the branching fraction i.e., as
\begin{align}
\mu^{\text{exp}}_W &= \sigma^{\text{exp}}(pp\to h X) \times \text{BR}^{\text{exp}}(h\to WW^* \to l' \nu' l\nu) \, \, \notag \\ 
\mu^{\text{exp}}_Z &= \sigma^{\text{exp}}(pp\to h X) \times \text{BR}^{\text{exp}}(h\to ZZ^* \to l'\, ^- l' \, ^+  l^- l^+) \, ,
\end{align}
where $\sigma(pp\to h X)$ stands for any specific production cross section. In the narrow width approximation we can make the following split
\begin{align}
\mu^{\text{exp}}_W &= \sigma^{\text{exp}}(pp\to h X) \times \text{BR}^{\text{exp}}(h\to W l\nu) \times \text{BR}^{\text{exp}}(W\to l' \nu')  \, \notag \\ 
\mu^{\text{exp}}_Z &= \sigma^{\text{exp}}(pp\to h X) \times \text{BR}^{\text{exp}}(h\to Z l^- l^+) \times \text{BR}^{\text{exp}}(Z\to l'\, ^- l' \, ^+) \, ,
\end{align}
In order to reduce the uncertainties, one can normalize the previous experimental signal strengths to the SM. As the experimentally measured decay width of the massive gauge bosons are in excellent agreement with the SM predictions, the normalized signal strength simplifies to
\begin{align}
\hat{\mu}_W &\equiv \frac{\mu^{\text{exp}}_W}{\mu^\text{SM}_W} = \frac{\sigma^{\text{exp}}(pp\to h X) \times \text{BR}^{\text{exp}}(h\to W l\nu)}{\sigma^{\text{SM}}(pp\to h X) \times \text{BR}^{\text{SM}}(h\to W l\nu)} \, , \notag \\[2ex]
\hat{\mu}_Z &\equiv \frac{\mu^{\text{exp}}_W}{\mu^\text{SM}_W} = \frac{\sigma^{\text{exp}}(pp\to h X) \times \text{BR}^{\text{exp}}(h\to Z l^- l^+)}{\sigma^{\text{SM}}(pp\to h X) \times \text{BR}^{\text{SM}}(h\to Z l^- l^+) } \, .
\end{align}
In order to compare the experimental results with the ATHDM predictions (as the model does not modify the massive gauge bosons decay widths at tree level, and the loop corrections are highly suppressed) we define the following generalized signal strengths
\begin{align}
{\mu}_W(\varphi_i^0) &\equiv \frac{\sigma^{\text{exp}}(pp\to \varphi_i^0 X) \times \text{BR}(\varphi_i^0 \to W l\nu)}{\sigma^{\text{SM}}(pp\to h X) \times \text{BR}^{\text{SM}}(h\to W l\nu)} \, , \notag \\[2ex]
{\mu}_Z (\varphi_i^0) &\equiv \frac{\sigma(pp\to \varphi_i^0 X) \times \text{BR}(\varphi_i^0 \to Z l^- l^+)}{\sigma^{\text{SM}}(pp\to h X) \times \text{BR}^{\text{SM}}(h\to Z l^- l^+) } \,  , \label{mu_signal}
\end{align}
that will be used for the fit to the LHC experimental data in our analysis.

\section{Integral Cancellations}
\label{AppInt}

Previously we have presented the cancellation of the integrals (\ref{INT}) in the form they appear in the expressions of $a^Z_\Lambda(q^2,\cos\theta)$, $b^Z_\Lambda(q^2,\cos\theta)$ and $c^Z_\Lambda(q^2,\cos\theta)$, for simplicity, for an easy interpretation of the corresponding cancellations. One should be easily able to check the integral cancellations using the following results. First, one should note that
\begin{align}
M_{\varphi_i^0}^2 - q^2 - 2p\cdot p_2 &= \frac{1}{2} ( M_{\varphi_i^0}^2 + M_Z^2 -q^2 ) \notag \\
& \qquad\qquad- \frac{1}{2q^2}\lambda^{1/2} (M_{\varphi_i^0}^2,M_W^2,q^2) \lambda^{1/2} (q^2,m_l^2,m_{l}^2) \cos \theta \, ,
\end{align}
and also
\begin{align}
M_Z^2 + 2p\cdot p_2 &= \frac{1}{2} ( M_{\varphi_i^0}^2 + M_Z^2 -q^2 ) \notag \\
& \qquad\qquad + \frac{1}{2q^2}\lambda^{1/2} (M_{\varphi_i^0}^2,M_W^2,q^2) \lambda^{1/2} (q^2,m_l^2,m_{l}^2) \cos \theta \, .
\end{align}
which are the expressions for the propagator denominators.  Therefore, all the given integrals are straightforward i.e., of the type
\begin{align}
I(a,b,m,m') = \int_{a}^b d\cos\theta \; \frac{(\cos\theta)^m}{(A \pm B\cos\theta)^{m'}} \, ,
\end{align}
or
\begin{align}
J(a,b,m'') = \int_{a}^b d\cos\theta \; \frac{(\cos\theta)^{m''}}{(A+B\cos\theta)(A-B\cos\theta)} \, ,
\end{align}
with the integration limits $(a,\,b)=\{(-1,0),\,(0,\, 1)\}$ and for $m=\{0,\,1,\,2\}$, $m'=\{1,\,2\}$ and $m''=\{0,\,2\}$, with $A,B\in \mathbb{R}$, $A,B > 0$, $A>B$. The cancellations follow straightforwardly.

\section{CP-violating scalar potential and real Yuakawa coupling}
\label{CP-pot}

As previously analysed in~\cite{Celis:2013rcs}, considering small imaginary parts of the $\lambda_{5,6}$ parameters of the potential i.e., $\lambda_{5,6}^I\ll$, we can write the scalar field mixing given by the $\mathcal{R}$ matrix as
\begin{align}
\begin{pmatrix}
h \\ H \\ A 
\end{pmatrix} 
=
\begin{pmatrix}
 \cos\tilde{\alpha} & \sin\tilde{\alpha} & \epsilon_{13} \\
-\sin\tilde{\alpha} & \cos\tilde{\alpha} & \epsilon_{23} \\
             \epsilon_{31}      & \epsilon_{32}         & 1
\end{pmatrix} 
\begin{pmatrix}
S_1 \\ S_2 \\ S_3 
\end{pmatrix} 
\, ,
\end{align}
where $\epsilon_{ij}$ are small parameters that further depend on the scalar boson masses and the $\lambda_i$ parameters of the potential. Their explicit expressions can be found
 in~\cite{Celis:2013rcs}.

As we have previously seen, the total decay rates are not affected by the extra contributions, therefore the potentially measurable effects are given by the numerator of (\ref{A_lambda}). For $V=W$ the only term that contains CP-violating effects is given by Im$(y^h_{l,d})$ which will given by 
\begin{align}
\text{Im}(y^h_{l,d}) = \epsilon_{13} \, \varsigma_{l,d} \, . 
\end{align}
As for $V=Z$, due to the fact that the neutral scalars do not have a definite CP quantum number but Re$(y^A_f)=0$, diagram $(2)$ will only contain contributions from $H$ (note the difference with the case where we had CP violation in the Yukawa sector, where the contribution was from $A$). In expression (\ref{B12}) for example, we have (for $\varphi_{i=1}^0 = h$)
\begin{align}
\sum_k  \frac{1}{q^2 - M_{\varphi_k^0}^2} (\mathcal{R}_{13}\mathcal{R}_{k2} - \mathcal{R}_{12}\mathcal{R}_{k3}) \, \text{Re} \left( y_{l}^{\varphi_k^0} \right) 
& = \frac{- c\tilde{\alpha} \, s\tilde{\alpha} \, (\epsilon_{13} + \varsigma_l \, \epsilon_{23}) + \varsigma_l \, \epsilon_{13} \, c^2\tilde{\alpha} + s^2\tilde{\alpha} \, \epsilon_{23}}{q^2 - {M}_H^2} \, ,
\end{align}  
where we have introduced the short-hand notation $c\tilde{\alpha} \equiv \cos\tilde{\alpha}$ and $s\tilde{\alpha} \equiv \sin\tilde{\alpha}$. Similarly, the only surviving term from $\mathcal{G}_{ij}$ (again, for $i=1$) in (\ref{Gij}) is given by
\begin{align}
\mathcal{G}_{12} = \frac{c\tilde{\alpha} \, \epsilon_{13} - s\tilde{\alpha} \, \epsilon_{23}}{q^2 - {M}_H^2} \, .
\end{align}
The remaining relevant terms are $\mathcal{Y}^\text{I}_{12} = 0$ and $\mathcal{Y}^\text{R}_{12}$ which is given by the usual expression
\begin{align}
\mathcal{Y}^\text{R}_{12} = (\cos\tilde{\alpha} + \varsigma_{l,d} \sin\tilde{\alpha}) (-\sin\tilde{\alpha} + \varsigma_{l,d} \cos\tilde{\alpha})  \, .
\end{align}

\section*{Acknowledgements}

I would like to thank Antonio Pich for helpful comments and discussions on this manuscript.

\end{appendix}



\begin{thebibliography}{100}





\bibitem{Aad:2012tfa}
{\scshape ATLAS} collaboration, G.~Aad et~al., \emph{{Observation of a new
  particle in the search for the Standard Model Higgs boson with the ATLAS
  detector at the LHC}},
  {\emph{Phys. Lett.}
  {\bf B716} (2012) 1--29}, 


\bibitem{Chatrchyan:2012xdj}
{\scshape CMS} collaboration, S.~Chatrchyan et~al., \emph{{Observation of a new
  boson at a mass of 125 GeV with the CMS experiment at the LHC}},
  {\emph{Phys. Lett.}
  {\bf B716} (2012) 30--61}, 



\bibitem{Aad:2019mbh}
{\scshape ATLAS} collaboration, G.~Aad et~al., \emph{{Combined measurements of
  Higgs boson production and decay using up to $80$ fb$^{-1}$ of proton-proton
  collision data at $\sqrt{s}=$ 13 TeV collected with the ATLAS experiment}},
  {\emph{Phys. Rev.}
  {\bfseries D101} (2020) 012002},




\bibitem{Cadamuro:2019tcf}
{\scshape ATLAS, CMS} collaboration, L.~Cadamuro, \emph{{Higgs boson couplings
  and properties}}, 
  {\emph{PoS}
  {\bfseries LHCP2019} (2019) 101}.




\bibitem{ATLAS-CONF-2019-004}
{\scshape ATLAS} collaboration, \emph{{Measurement of Higgs boson production in
  association with a $t\overline t$ pair in the diphoton decay channel using
  139~fb$^{-1}$ of LHC data collected at $\sqrt{s} = 13$~TeV by the ATLAS
  experiment}},  Tech. Rep. ATLAS-CONF-2019-004, CERN, Geneva, Mar, 2019.



\bibitem{Aaboud:2018urx}
{\scshape ATLAS} collaboration, M.~Aaboud et~al., \emph{{Observation of Higgs
  boson production in association with a top quark pair at the LHC with the
  ATLAS detector}},
  {\emph{Phys. Lett.}
  {\bfseries B784} (2018) 173--191},




\bibitem{CMS-PAS-HIG-19-005}
{\scshape CMS} collaboration, \emph{{Combined Higgs boson production and decay
  measurements with up to 137 fb-1 of proton-proton collision data at $\sqrt{s}
  = 13$ TeV}},  Tech. Rep. CMS-PAS-HIG-19-005, CERN, Geneva, 2020.



\bibitem{Sirunyan:2018koj}
{\scshape CMS} collaboration, A.~M. Sirunyan et~al., \emph{{Combined
  measurements of Higgs boson couplings in proton-proton collisions at
  $\sqrt{s}=13~$ TeV}},
  {\emph{Eur. Phys. J.}
  {\bfseries C79} (2019) 421},



\bibitem{Sirunyan:2018kst}
{\scshape CMS} collaboration, A.~M. Sirunyan et~al., \emph{{Observation of
  Higgs boson decay to bottom quarks}},
  {\emph{Phys. Rev. Lett.}
  {\bfseries 121} (2018) 121801},




\bibitem{Aaboud:2018jqu}
M.~Aaboud \textit{et al.} [ATLAS],
\emph{{Measurements of gluon-gluon fusion and vector-boson fusion Higgs boson production cross-sections in the $H \to WW^{\ast} \to e\nu\mu\nu$ decay channel in $pp$ collisions at $\sqrt{s}=13$ TeV with the ATLAS detector}} 
Phys. Lett. B \textbf{789} (2019), 508-529



\bibitem{Aad:2019lpq}
G.~Aad \textit{et al.} [ATLAS],
emph{{Measurement of the production cross section for a Higgs boson in association with a vector boson in the $H \to WW^{\ast} \to \ell\nu\ell\nu$ channel in $pp$ collisions at $\sqrt{s}$ = 13 TeV with the ATLAS detector}} 
Phys. Lett. B \textbf{798} (2019), 134949


\bibitem{Sirunyan:2020tzo}
A.~M.~Sirunyan \textit{et al.} [CMS],
\emph{{Measurement of the inclusive and differential Higgs boson production cross sections in the leptonic WW decay mode at $\sqrt{s} =$ 13 TeV}}


\bibitem{Sirunyan:2017exp}
A.~M.~Sirunyan \textit{et al.} [CMS],
\emph{{Measurements of properties of the Higgs boson decaying into the four-lepton final state in pp collisions at $ \sqrt{s}=13 $ TeV}}
JHEP \textbf{11} (2017), 047



\bibitem{Sirunyan:2018sgc}
A.~M.~Sirunyan \textit{et al.} [CMS],
\emph{{Measurement and interpretation of differential cross sections for Higgs boson production at $\sqrt{s} =$ 13 TeV}}
Phys. Lett. B \textbf{792} (2019), 369-396




\bibitem{ATLAS:2020wny}
G.~Aad \textit{et al.} [ATLAS],
\emph{{Measurements of the Higgs boson inclusive and differential fiducial cross sections in the 4$\ell$ decay channel at $\sqrt{s}$ = 13 TeV}}
Eur. Phys. J. C \textbf{80} (2020) no.10, 942




\bibitem{Aaboud:2017oem}
M.~Aaboud \textit{et al.} [ATLAS],
\emph{{Measurement of inclusive and differential cross sections in the $H \rightarrow ZZ^* \rightarrow 4\ell$ decay channel in $pp$ collisions at $\sqrt{s}=13$ TeV with the ATLAS detector}} 
JHEP \textbf{10} (2017), 132



\bibitem{Sirunyan:2017tqd}
A.~M.~Sirunyan \textit{et al.} [CMS],
\emph{{Constraints on anomalous Higgs boson couplings using production and decay information in the four-lepton final state}} 
Phys. Lett. B \textbf{775} (2017), 1-24



\bibitem{Bhattacharya:2020lfm}
B.~Bhattacharya, A.~Datta, S.~Kamali and D.~London, \emph{{A measurable angular distribution for $ \overline{B}\to {D}^{\ast }{\tau}^{-}{\overline{v}}_{\tau } $ decays}}
JHEP \textbf{07} (2020) no.07, 194


\bibitem{Asadi:2020fdo} P.~Asadi, A.~Hallin, J.~Martin Camalich, D.~Shih and S.~Westhoff, \emph{{Complete framework for tau polarimetry in $B\to D^{(*)}\tau\nu$  decays}}
Phys. Rev. D \textbf{102} (2020) no.9, 095028

\bibitem{Hill:2019zja}
D.~Hill, M.~John, W.~Ke and A.~Poluektov, \emph{{Model-independent method for measuring the angular coefficients of $B^0 \to D^{*-} \tau^+ \nu_{\tau}$ decays}}
JHEP \textbf{11} (2019), 133
 

\bibitem{Tanaka:1994ay}
M.~Tanaka, \emph{{Charged Higgs effects on exclusive semitauonic $B$ decays}}
Z. Phys. C \textbf{67} (1995), 321-326


\bibitem{Sakaki:2013bfa} 
Y.~Sakaki, M.~Tanaka, A.~Tayduganov and R.~Watanabe, \emph{{Testing leptoquark models in $\bar B \to D^{(*)} \tau \bar\nu$}}
Phys. Rev. D \textbf{88} (2013) no.9, 094012

  
\bibitem{KEUNE} 
Anne Keune, \emph{{Reconstruction of the Tau Lepton and the Study of $B^0 \to D^{*-} \tau^+ \bar\nu$´ at LHCb}}
Th\`ese NO 5384 (2012), \'Ecole Polytechnique F\'ed\'erale de Lausanne








\bibitem{Datta:2012qk} 
A.~Datta, M.~Duraisamy and D.~Ghosh, \emph{{Diagnosing New Physics in $b \to c \, \tau \, \nu_\tau$ decays in the light of the recent BaBar result}}
Phys. Rev. D \textbf{86} (2012), 034027 


\bibitem{Celis:2012dk}
A.~Celis, M.~Jung, X.~Q.~Li and A.~Pich,
\emph{{Sensitivity to charged scalars in $\boldsymbol{B\to D^{(*)}\tau\nu_\tau}$ and $\boldsymbol{B\to\tau\nu_\tau}$ decays}}
JHEP \textbf{01} (2013), 054


\bibitem{Fajfer:2012vx}
S.~Fajfer, J.~F.~Kamenik and I.~Nisandzic, \emph{{On the $B \to D^* \tau \bar \nu_{\tau}$ Sensitivity to New Physics}}
Phys. Rev. D \textbf{85} (2012), 094025


\bibitem{Hu:2018veh}
Q.~Y.~Hu, X.~Q.~Li and Y.~D.~Yang, \emph{{$b\to c\tau\nu$ transitions in the standard model effective field theory}}
Eur. Phys. J. C \textbf{79} (2019) no.3, 264


\bibitem{Murgui:2019czp} 
C.~Murgui, A.~Pe\~nuelas, M.~Jung and A.~Pich, \emph{{Global fit to $b \to c \tau \nu$ transitions}}
JHEP \textbf{09} (2019), 103

\bibitem{Alonso:2017ktd}
R.~Alonso, J.~Martin Camalich and S.~Westhoff, \emph{{Tau properties in $B\to D\tau\nu$ from visible final-state kinematics}}
Phys. Rev. D \textbf{95} (2017) no.9, 093006


\bibitem{Cheung:2020sbq}
K.~Cheung, Z.~R.~Huang, H.~D.~Li, C.~D.~L\"u, Y.~N.~Mao and R.~Y.~Tang, \emph{{Revisit to the $b\to c\tau\nu$ transition: In and beyond the SM}}
Nucl. Phys. B \textbf{965} (2021), 115354





\bibitem{Gonzalez-Alonso:2014eva}
M.~Gonzalez-Alonso, A.~Greljo, G.~Isidori and D.~Marzocca,
\emph{{Pseudo-observables in Higgs decays}} 
Eur. Phys. J. C \textbf{75} (2015), 128


\bibitem{Greljo:2015sla}
A.~Greljo, G.~Isidori, J.~M.~Lindert and D.~Marzocca,
\emph{{Pseudo-observables in electroweak Higgs production}}
Eur. Phys. J. C \textbf{76} (2016) no.3, 158





\bibitem{Isidori:2013cla}
G.~Isidori, A.~V.~Manohar and M.~Trott,
\emph{{Probing the nature of the Higgs-like Boson via $h \to V \mathcal{F}$ decays}}
Phys. Lett. B \textbf{728} (2014), 131-135


\bibitem{Pomaroll}
R.S.~Gupta, A.~Pomarol and F.~Riva,
\emph{{Leading effects beyond the standard model}}
Phys. Rev. D \textbf{91} (2015), 035001


\bibitem{Banerjee:2019pks}
S.~Banerjee, R.~S.~Gupta, J.~Y.~Reiness and M.~Spannowsky,
\emph{{Resolving the tensor structure of the Higgs coupling to $Z$-bosons via Higgs-strahlung}}
Phys. Rev. D \textbf{100} (2019) no.11, 115004


\bibitem{Falkowski:2015jaa}
A.~Falkowski, M.~Gonzalez-Alonso, A.~Greljo and D.~Marzocca,
\emph{{Global constraints on anomalous triple gauge couplings in effective field theory approach}}
Phys. Rev. Lett. \textbf{116} (2016) no.1, 011801










\bibitem{Eberhardt:2020dat}
O.~Eberhardt, A.~P.~Mart\'\i{}nez and A.~Pich,
\emph{{Global fits in the Aligned Two-Higgs-Doublet model}} 
[arXiv:2012.09200 [hep-ph]].



\bibitem{Abbas:2018pfp}
G.~Abbas, D.~Das and M.~Patra,
\emph{{Loop induced $H^\pm \to W^\pm Z$ decays in the aligned two-Higgs-doublet model}}
Phys. Rev. D \textbf{98} (2018) no.11, 115013


\bibitem{Hu:2016gpe}
Q.~Y.~Hu, X.~Q.~Li and Y.~D.~Yang,
\emph{{$B^0\to K^{\ast 0}\mu^+\mu^-$ decay in the Aligned Two-Higgs-Doublet Model}}
Eur. Phys. J. C \textbf{77} (2017) no.3, 190







\bibitem{Abbas:2015cua}
G.~Abbas, A.~Celis, X.~Q.~Li, J.~Lu and A.~Pich,
\emph{{Flavour-changing top decays in the aligned two-Higgs-doublet model}}
JHEP \textbf{06} (2015), 005



\bibitem{Ilisie:2015tra}
V.~Ilisie,
\emph{{New Barr-Zee contributions to $\mathbf{(g-2)_\mu}$ in two-Higgs-doublet models}}
JHEP \textbf{04} (2015), 077




\bibitem{Celis:2013ixa}
A.~Celis, V.~Ilisie and A.~Pich,
\emph{{Towards a general analysis of LHC data within two-Higgs-doublet models}}
JHEP \textbf{12} (2013), 095





\bibitem{Jung:2010ab}
M.~Jung, A.~Pich and P.~Tuzon,
\emph{{The $B \to X_s$ gamma Rate and CP Asymmetry within the Aligned Two-Higgs-Doublet Model}}
Phys. Rev. D \textbf{83} (2011), 074011

  
  
\bibitem{Jung:2010ik}
M.~Jung, A.~Pich and P.~Tuzon,
\emph{{Charged-Higgs phenomenology in the Aligned two-Higgs-doublet model}}
JHEP \textbf{11} (2010), 003




\bibitem{Celis:2013jha}
A.~Celis, M.~Jung, X.~Q.~Li and A.~Pich,
\emph{{$B \to D^* \tau \nu \tau$ decays in two-Higgs-doublet models}}
J. Phys. Conf. Ser. \textbf{447} (2013), 012058



\bibitem{Jung:2013hka}
M.~Jung and A.~Pich,
\emph{{Electric Dipole Moments in Two-Higgs-Doublet Models}}
JHEP \textbf{04} (2014), 076

























\bibitem{Celis:2013rcs}
A.~Celis, V.~Ilisie and A.~Pich,
\emph{{LHC constraints on two-Higgs doublet models}}
JHEP \textbf{07} (2013), 053



\bibitem{Botella:2015hoa}
F.~J.~Botella, G.~C.~Branco, M.~Nebot and M.~N.~Rebelo,
\emph{{Flavour Changing Higgs Couplings in a Class of Two Higgs Doublet Models}}
Eur. Phys. J. C \textbf{76} (2016) no.3, 161


\bibitem{Glashow:1976nt}
S.~L.~Glashow and S.~Weinberg,
\emph{{Natural Conservation Laws for Neutral Currents}}
Phys. Rev. D \textbf{15} (1977), 1958



\bibitem{Pich:2009sp}
A.~Pich and P.~Tuzon,
\emph{{Yukawa Alignment in the Two-Higgs-Doublet Model}}
Phys. Rev. D \textbf{80} (2009), 091702







\bibitem{Keung}
Keung, Wai-Yee and Marciano, William J.
\emph{{Higgs-scalar decays: $H\ensuremath{\rightarrow}{W}^{\ifmmode\pm\else\textpm\fi{}}+X$}},
Phys. Rev. D \textbf{30}, 1 (1984), 248--250



\bibitem{Djouadi:2005gi}
A.~Djouadi,
\emph{{The Anatomy of electro-weak symmetry breaking. I: The Higgs boson in the standard model}},
Phys. Rept. \textbf{457} (2008), 1-216














\end{thebibliography}
\end{document}